\documentclass[journal]{IEEEtran}
\usepackage{cite,units}
\usepackage{wrapfig}
\ifCLASSINFOpdf
   \usepackage[pdftex]{graphicx}
  \DeclareGraphicsExtensions{.pdf,.jpeg,.png,.eps}
\else
  \usepackage[dvips]{graphicx}
   \graphicspath{{../eps/}}
  \DeclareGraphicsExtensions{.eps}
\fi
\usepackage{amsfonts,amssymb}
\usepackage{array}
\usepackage[tight,footnotesize]{subfigure}
\usepackage[cmex10]{amsmath}
\usepackage{cleveref}
\usepackage{mdwmath}
\usepackage{mdwtab}

\usepackage{float}

\crefname{equation}{equation}{equations}
\Crefname{equation}{Equation}{Equations}
\crefrangelabelformat{equation}{(#3#1#4--#5#2#6)}
\crefmultiformat{equation}{equations (#2#1#3}{, #2#1#3)}{#2#1#3}{#2#1#3}
\Crefmultiformat{equation}{Equations (#2#1#3}{, #2#1#3)}{#2#1#3}{#2#1#3}
\allowdisplaybreaks[2]
\begin{document}
\title{Experimental Verification of Below-Cutoff Propagation in Miniaturized Circular Waveguides Using Anisotropic ENNZ Metamaterial Liners}
\author{Justin~G.~Pollock,~\IEEEmembership{Student~Member,~IEEE,}
        and~Ashwin~K.~Iyer,~\IEEEmembership{Senior Member,~IEEE,}%
}
\maketitle
\begin{abstract}
This paper presents experimental verification of below-cutoff transmission through miniaturized waveguides whose interior is coated with a thin anisotropic metamaterial liner possessing epsilon-negative and near-zero (ENNZ) properties. These liners are realized using a simple, printed-circuit implementation based on inductively loaded wires, and introduce an {\em HE}$_{11}$ mode well below the natural cutoff frequency. The inclusion of the liner is shown to substantially improve the transmission between two embedded shielded-loop sources. A homogenization scheme is developed to characterize the liner's anisotropic effective-medium parameters, which is shown to accurately describe a set of frequency-reduced cutoffs. The fabrication of the lined waveguide is discussed, and the experimental and simulated transmission results are shown to be in agreement.
\end{abstract}

\section{Introduction}
\label{sec:intro}
\IEEEPARstart{M}ETALLIC waveguides are attractive at high frequencies for their low propagation loss, moderate bandwidths, shielding properties, and excellent power-handling capability. Hollow versions prove useful in several applications, including fluid-density measurements in the oil/gas industry~\cite{penirschke2008microwave}, microwave heating of fluids~\cite{ratanadecho2002numerical}, the observation of exotic radiation in particle physics~\cite{yin1993cyclotron}, and low-loss infrared transmission of CO$_2$-laser light~\cite{nubling1996hollow}. However, these waveguides lack the capability to be miniaturized without additional loading, due to the strict dependence of their cutoff frequencies on their transverse dimensions. Inhomogeneously filling the waveguide with dielectric regions relaxes this condition, retains access to their interiors, and can introduce new propagation phenomena, such as frequency-reduced propagating bands with forward-wave or backward-wave behaviour~\cite{clarricoats1964evanescent}.

With the advent of metamaterial loading of waveguides, it is possible to obtain intriguing propagation phenomena through engineering the metamaterial's effective permittivity and permeability. For instance, below-cutoff propagation occurs in waveguides loaded by arrays of subwavelength electric and magnetic scatters that exhibit negative permittivity and/or permeability at low frequencies~\cite{hrabar2005waveguide,meng2011controllable}. It was recently shown that the inclusion of thin metamaterial liners into PEC circular waveguides can permit propagation well below the unlined waveguide's fundamental-mode cutoff~\cite{pollockmttt2013}. Modeling the liner as isotropic and homogeneous, a frequency-reduced backward-wave band corresponding to a hybrid electric {\em HE}$_{11}$ mode was shown to occur in the frequency regime in which the liner exhibited dispersive epsilon-negative and near-zero (ENNZ) properties. This backward-wave band was over $42\%$ below the waveguide's fundamental-mode cutoff frequency, in which transmission occurred through multiple resonant peaks corresponding to Fabry-P\'erot-type resonant phase conditions over the waveguide's length. Although narrowband and dispersive, it was suggested in~\cite{pollockmttt2013} that these waveguides may enable several emerging applications, including traveling-wave imaging in MRI scanner bores~\cite{PollockISMRM2012}, characterization of the properties of small quantities of fluids~\cite{alu2008dielectric}, and in novel particle-beam experiments~\cite{duan2009research} -- all of which require access to the waveguide interior and can benefit greatly from miniaturization. Complementing these transmission studies, the frequency-reduced {\em HE}$_{11}$ mode in a metamaterial-lined open-ended waveguide (OEWG) probe antenna was shown in simulation to yield substantial gain improvements over a similarly sized hollow and below-cutoff waveguide probe~\cite{pollockTAP2014}. In each of these works, the subwavelength waveguide dimensions and the stark contrast between the metamaterial and vacuum permittivities produced an {\em HE}$_{11}$-mode field profile with uniform fields in the inner-vacuum region and discontinuously strong fields in the outer-liner region.

While these initial studies examined the instructive case in which the liner is modeled as an effective medium with an isotropic and homogeneous permittivity, practical metamaterials are naturally anisotropic. For instance, optical fibers loaded using radially emanating silver-coated nanopores~\cite{pollockopticsx2015} exhibit plasmonic-like behaviour. In these waveguides, the metamaterial's permittivity is anisotropic and may be modeled by a diagonalized tensor ($\bar{\bar{\epsilon}}$) in a cylindrical coordinate system with elements that assume negative values for the directions corresponding to the nanopore axes. In the microwave regime, these negative permittivities may be achieved using arrays of inductively loaded thin wires, which can be implemented in printed-circuit form~\cite{pollockAEC2013,pollockTAP2014}. Although homogenization of such metamaterials can be carried out through field averaging~\cite{smith2006homogenization} or inversion of the scattering properties of a metamaterial sample~\cite{smith2005electromagnetic}, these approaches have difficulty modeling strong spatial dispersion~\cite{elser2007nonlocal}, bianisotropy~\cite{alu2011first}, and non-TEM propagation, all of which may be observed in thin-wire media. This is further complicated in cylindrical geometries. An approach for the accurate homogenization of cylindrically anisotropic metamaterials might bring significant clarity to intriguing phenomena recently observed in metamaterial-loaded waveguides~\cite{duan2009research}.

The layout of the paper is as follows: Section~\ref{sec:theory} explores the dispersion of several new modes supported by metamaterial-lined circular waveguides whose liner is modeled as an anisotropic and homogeneous effective medium, developing relationships between individual material parameters and the cutoff frequencies of modes of interest. Section~\ref{sec:real} develops a printed-circuit implementation of the ENNZ metamaterial liner and introduces a novel homogenization approach that gives insight into the metamaterial's effective-medium parameters based on the dispersions of frequency-reduced modes supported by the liner and their similarity to the transmission-line (TL) modes of TL metamaterials. An experimental prototype is developed, and parametric studies are performed to gauge the impact of fabrication tolerances on the dispersion of frequency-reduced modes. In Sec.~\ref{sec:full}, the metamaterial-lined waveguide is placed between two embedded shielded loops, and the transmission features are investigated through full-wave simulations and experiments.

\section{Theory}
\label{sec:theory}
\begin{figure}[!t]
\centering
\includegraphics[width=3in]{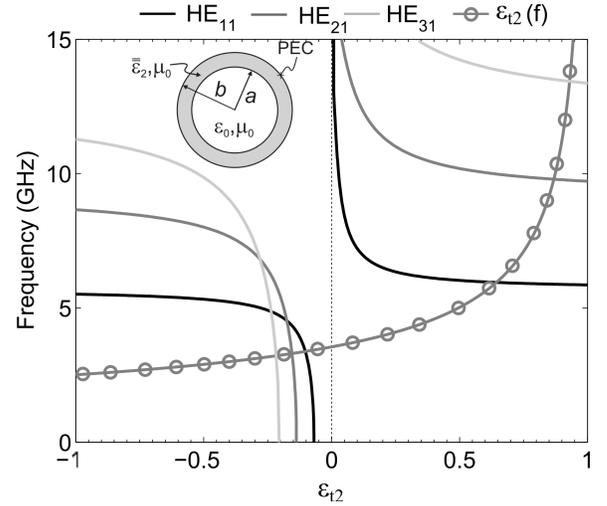}
\caption{\unskip Cutoff frequencies of the {\em HE}$_{11}$, {\em HE}$_{21}$, {\em HE}$_{31}$ modes versus the relative permittivity, $\epsilon_{t2}$, of the anisotropic dielectric liner $\left[\bar{\bar{\epsilon}}_2=(\epsilon_{t2},\epsilon_{t2},1)\epsilon_0\right]$ of the host PEC circular waveguide. The dispersion profile of $\epsilon_{t2}(f)$ defined in Sec.~\ref{sec:theory} is shown.\label{fig1}}
\end{figure}

The inset in Fig.~\ref{fig1} presents the geometry of the metamaterial-lined circular waveguide (radius $b$) under consideration. An inner-core vacuum region (radius $a$) of permittivity $\epsilon_0$ and permeability $\mu_0$ is surrounded by a metamaterial layer of thickness $t=b-a$, backed by a PEC waveguide wall. In contrast to~\cite{pollockmttt2013}, we investigate here the more complex, but more realistic, case in which the metamaterial liner has an anisotropic permittivity and permeability described by tensors diagonalized in a cylindrical coordinate system. Taking the coordinate axis to coincide with the waveguide axis, the material tensors take the form $\bar{\bar{\epsilon}}_2=I(\epsilon_{\rho2},\epsilon_{\phi2},\epsilon_{z2})\epsilon_0$ and $\bar{\bar{\mu}}_2=I(\mu_{\rho2},\mu_{\phi2},\mu_{z2})\mu_0$, in which $I$ is the identity tensor. After substantial mathematical manipulation, the resulting anisotropic metamaterial-lined circular waveguide's dispersion relation takes the following form:
\begin{subequations}
\begin{align}
AB&=\frac{-n^2\gamma^2}{k^2_0 a^2}\left[\frac{\epsilon_{z2}}{\epsilon_{\rho2}{(\gamma^{\epsilon}_{\rho2})}^2} - \frac{1}{\gamma^2_{{\rho1}}}\right]\left[\frac{\mu_{z2}}{\mu_{\rho2}{(\gamma^{\mu}_{\rho2})}^2} - \frac{1}{\gamma^2_{{\rho1}}}\right]\label{EQ1a}\\
A&=\left[\frac{1}{\gamma_{{\rho1}}}\frac{J'_{n}(\gamma_{\rho1}a)}{J_{n}(\gamma_{\rho1}a)}-\frac{\mu_{z2}}
{\gamma^{\mu}_{\rho2}}\frac{G'_{\nu_{\mu}}(\gamma^{\mu}_{\rho2}a)}{G_{\nu_{\mu}}(\gamma^{\mu}_{\rho2}a)}\right]\label{EQ1b}\\
B&=\left[\frac{1}{\gamma_{\rho1}}
\frac{J'_{n}(\gamma_{\rho1}a)}{J_{n}(\gamma_{\rho1}a)}-\frac{\epsilon_{z2}}{\gamma^{\epsilon}_{\rho2}}
\frac{F'_{\nu_{\epsilon}}(\gamma^{\epsilon}_{\rho2}a)}{F_{\nu_{\epsilon}}(\gamma^{\epsilon}_{\rho2}a)}\right]\label{EQ1c}
\end{align}
\end{subequations}
Here $\gamma=\alpha+j\beta$ is the complex waveguide propagation constant in the $z$ (axial) direction, $k_0=\omega\sqrt{\epsilon_0\mu_0}$, $\gamma_{\rho1}=\sqrt{\gamma^2+k_0^2}$, and we define the following quantities:
\begin{subequations}
\begin{align}
\gamma^{\epsilon}_{\rho2}&=\sqrt{\frac{\epsilon_{z2}}{\epsilon_{\rho2}}}\sqrt{\gamma^2+{k_0}^2\epsilon_{\rho2}\mu_{\phi2}}\label{EQ2a}\\
\gamma^{\mu}_{\rho2}&=\sqrt{\frac{\mu_{z2}}{\mu_{\rho2}}}\sqrt{\gamma^2+{k_0}^2\epsilon_{\phi2}\mu_{\rho2}}\label{EQ2b}
\end{align}
\end{subequations}
$J_n$ is a Bessel function of integer order $n$, and $F_{\nu_{\epsilon}}$ and $G_{\nu_{\mu}}$ are combinations of Bessel ($J_{\nu}$) and Neumann ($Y_{\nu}$) functions, which can be expressed as follows:
\begin{subequations}
\begin{align}
F_{\nu_{\epsilon}}(\gamma^{\epsilon}_{\rho2}\rho)&=Y_{\nu_{\epsilon}}(\gamma^{\epsilon}_{\rho2}b)J_{\nu_{\epsilon}}(\gamma^{\epsilon}_{\rho2}\rho)-J_{\nu_{\epsilon}}(\gamma^{\epsilon}_{\rho2}b)Y_{\nu_{\epsilon}}(\gamma^{\epsilon}_{\rho2}\rho)\label{EQ3a}\\
G_{\nu_{\mu}}(\gamma^{\mu}_{\rho2}\rho)&=Y'_{\nu_{\mu}}(\gamma^{\mu}_{\rho2}b)J_{\nu_{\mu}}(\gamma^{\mu}_{\rho2}\rho)-J'_{\nu_{\mu}}(\gamma^{\mu}_{\rho2}b)Y_{\nu_{\mu}}(\gamma^{\mu}_{\rho2}\rho)\label{EQ3b}
\end{align}
\end{subequations}
Their orders are generally complex and are defined as follows:
\begin{subequations}
\begin{align}
\nu_{\epsilon}= \sqrt{\frac{\epsilon_{\phi2}\mu_{z2}}{\epsilon_{z2}\mu_{\rho2}}}\frac{\gamma^{\epsilon}_{\rho2}}{\gamma^{\mu}_{\rho2}}n, \;\;\;
\nu_{\mu}= \sqrt{\frac{\mu_{\phi2}\epsilon_{z2}}{\mu_{z2}\epsilon_{\rho2}}}\frac{\gamma^{\mu}_{\rho2}}{\gamma^{\epsilon}_{\rho2}}n\label{EQ4a}
\end{align}
\end{subequations}

The roots of Eqs.~(1b) and~(1c) respectively provide the cutoffs of the {\em HE} and {\em EH} modes. Whereas the resulting dispersion relation is capable of treating fully biaxial liners, the focus of this work is on {\em HE} modes, with field polarizations similar to their {\em TE} counterparts in a homogeneously filled waveguide with no longitudinal electric-field component at their cutoff frequencies. Hence, we look at several frequency-reduced {\em HE} modes that can be potentially supported by the simplifying case of a liner that has a biaxial permittivity (i.e., $\epsilon_{\rho2}\neq\epsilon_{\phi2}$) with $\epsilon_{z2}=1$ and a nonmagnetic response (i.e., $\mu_{\rho2}=\mu_{\phi2}=\mu_{z2}=1$). In this case, after applying the cutoff frequency condition (i.e., $\gamma=0$), Eqs.~(2b) and~(4a) reduce to $\gamma^{\mu}_{\rho2}=k_0(\epsilon_{\phi2})^{1/2}$ and $\nu_{\mu}=n(\epsilon_{\phi2}/\epsilon_{\rho2})^{1/2}$, respectively. When applied to Eq.~(1b), this reveals that the {\em HE} modes' cutoffs are independent of $\epsilon_{z2}$, but vary with $\epsilon_{\rho2}$ and $\epsilon_{\phi2}$. However, it should be noted from Eq.~(1b) that $\epsilon_{\rho2}$ is in the order of the Bessel functions while $\epsilon_{\phi2}$ is in the argument; thus, the {\em HE} modes' cutoff frequencies are far more sensitive to $\epsilon_{\rho2}$ than to $\epsilon_{\phi2}$. Due to this weak dependence on $\epsilon_{\phi2}$, for the remainder of the work we assume a uniaxial transverse permittivity $\epsilon_{t2}=\epsilon_{\rho2}=\epsilon_{\phi2}$. This assumption will aid in the theoretical analysis of the {\em HE} modes since the Bessel-function order becomes purely real.
\subsection{Cutoff-Frequency Dependence}
Consider a metamaterial-lined waveguide with dimensions $b=15$mm and $t=2.3$mm. Figure~\ref{fig1} presents the dependence of the dielectric-lined circular waveguide's {\em HE}$_{11}$ (black solid curve), {\em HE}$_{21}$ (dark grey solid curve) and {\em HE}$_{31}$ (light grey solid curve) modes' cutoff frequencies on $\epsilon_{t2}$. The cutoff frequencies are only marginally decreased for $|\epsilon_{t2}|>1$, which corroborates the result in~\cite{pollockmttt2013} for isotropic liners. As $|\epsilon_{t2}|\rightarrow0$, termed the epsilon-near-zero (ENZ) regime, the cutoff frequencies vary rapidly. For epsilon-positive and near-zero (EPNZ) values, all cutoff frequencies are strongly increased. However, as $\epsilon_{t2}\rightarrow 0^-$ (i.e., the ENNZ regime) the cutoff frequencies are drastically decreased. Equations~($1$) confirm all the above properties for all higher-azimuthal-order {\em HE}$_{n1}$ modes, which suggests that practical anisotropic metamaterial liners may introduce a useful diversity of frequency-reduced modes. In fact, as will be shown below for a particular ENNZ permittivity value, each mode's cutoff frequency can, in theory, be reduced to arbitrarily low frequencies for a particular ENNZ permittivity value. Furthermore, as the azimuthal order increases, the $\epsilon_{t2}$ required to effect this reduction tends to increasingly negative values. Therefore, for a sufficiently negative $\epsilon_{t2}$ the cutoff frequency of the {\em HE}$_{31}$ mode is lower than those of the {\em HE}$_{21}$ and {\em HE}$_{11}$ modes. This is opposite to the vacuum-filled case in which, as $n$ increases, the cutoff frequency of the {\em TE}$_{n1}$ mode increases.

To enable design of the {\em HE}$_{n1}$-mode cutoffs, we apply small-argument approximations to Eq.~(1b). This yields an approximate but useful relation between the {\em HE}$_{n1}$ cutoff frequency, $f_{c,n}$, and the waveguide's dimensions and liner properties:
\vspace{3pt}
\begin{subequations}
\begin{align}
f_{c,n}&=\left(\frac{2c}{2\pi a}\right)\sqrt{\frac{(n^2+n)(1-K)}{n-(n+2)K}}\;\; (n>0)\label{EQ5a}\\
\intertext{where}
 K&=\epsilon_{t2}\left[\frac{1+\left(\frac{b}{a}\right)^{2n}}{1-\left(\frac{b}{a}\right)^{2n}}\right]\;\;(n>0)\label{EQ5b}
\end{align}
\end{subequations}

This simple relationship constitutes a generalized design equation for all {\em HE}$_{n1}$ modes for $n>0$. The particular small-angle approximations that were used in deriving this expression are not valid for $n=0$, but the $f_c$ for these modes is most easily obtained numerically from Eqs.~($1$). By setting $K=1$, Eq.~(5a) captures the behaviour of the cutoff frequency tending to zero in Fig.~\ref{fig1}, which if applied to Eq.~(5b) yields:
\vspace{3pt}
\begin{equation}
\epsilon_{t2,max,n}=\frac{1-(b/a)^{2n}}{1+(b/a)^{2n}}.\label{EQ6}
\end{equation}
\\
In this expression, $\epsilon_{t2,max,n}$ specifies the maximum ENNZ value required by the anisotropic liner to arbitrarily lower the cutoff frequency, which is strictly dependent on the waveguide's and liner's dimensions. Equation~(6) explicitly captures the action of the ENNZ liners, in that as $a\rightarrow b$ (i.e., the liner is of infinitesimal thickness), $\epsilon_{t2,max,n}$, and therefore $\epsilon_{t2}$, must more closely approach zero to restore propagation. It should be noted that Eqs.~(5) become increasingly accurate both as $\epsilon_{t2}\rightarrow\epsilon_{t2,max,n}$ (hence, $f_c\rightarrow0$) and as $n$ increases. Furthermore, Eq.~(6) highlights the shift of $\epsilon_{t2,max,n}$ to more negative values as $n$ increases and in the limit $n\rightarrow\infty$, $\epsilon_{t2,max,n}\rightarrow -1$. This suggests a crowding of higher-azimuthal-order {\em HE}$_{n1}$ modes at $\epsilon_{t2}=-1$.

\subsection{Dispersion of Frequency-Reduced Modes}
Whereas Fig.~\ref{fig1} aids in choosing the value of $\epsilon_{t2}$ to yield a desired set of cutoff frequencies, the task of realizing ENNZ values requires that material dispersion for the liner be taken into account. For this reason, the liner's complex permittivity dispersion is described by a Drude model with $\epsilon_{t2}(f)=1-f_{ep}^2/(f(f-jf_t))$, in which $f_{ep}=3.958$GHz is the plasma frequency and $f_t=5$MHz is the damping frequency establishing the liner's loss.
\begin{figure}[!t]
\centering
\includegraphics[width=3.1in]{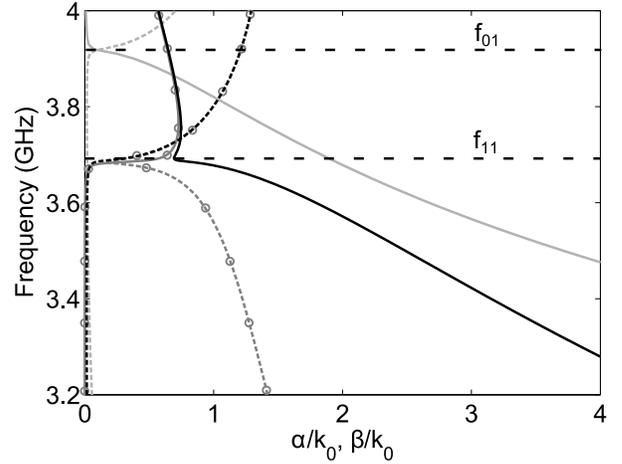}
\caption{\unskip Dispersion of $\alpha/k_0$ (dashed curves) and $\beta/k_0$ (solid curves) for the following modes of the metamaterial-lined waveguide: {\em EH}$_{01}$ (light grey curves), {\em HE}$_{11}$ waveguide-type (black curves), and {\em HE}$_{11}$ surface-type (dark grey circle curves). All curves are obtained from Eqs.~($1$).\label{fig2}}
\end{figure}

This $\epsilon_{t2}$ dispersion profile (dark grey circle curve) is overlayed on the cutoff-frequency curves in Fig.~\ref{fig1}. The intersections yield the corresponding {\em HE}$_{11}$, {\em HE}$_{21}$, and {\em HE}$_{31}$ cutoff frequencies, for which their frequency-reduced intersections occur for ENNZ $\epsilon_{t2}$ and the higher-frequency intersections occur near their natural cutoffs (since $\epsilon_{t2}$ approaches unity at these frequencies). Although Fig.~\ref{fig1} does not indicate the equivalent {\em EH} modes' intersections, in general the {\em EH} cutoff frequencies are weakly dependent on $\epsilon_{t2}$. Therefore, they would also occur at their natural cutoffs. However, for the {\em EH}$_{01}$ mode, $\epsilon_{t2}(f_{ep})=0$ satisfies Eq.~(1c); this corresponds to the intersection of the vertical axis with $\epsilon_{t2}(f)$ in Fig.~\ref{fig1}. The resulting important implication is the cutoff of this frequency-reduced {\em EH}$_{01}$ mode occurs precisely at the Drude plasma frequency of $\epsilon_{t2}$. This condition is independent of the dimensions and remaining material parameters, and occurs in addition to its high-frequency natural cutoff.
\begin{figure}[!t]
\centering
\subfigure[]{
\raisebox{0.25in}{\includegraphics[width=1.61in]{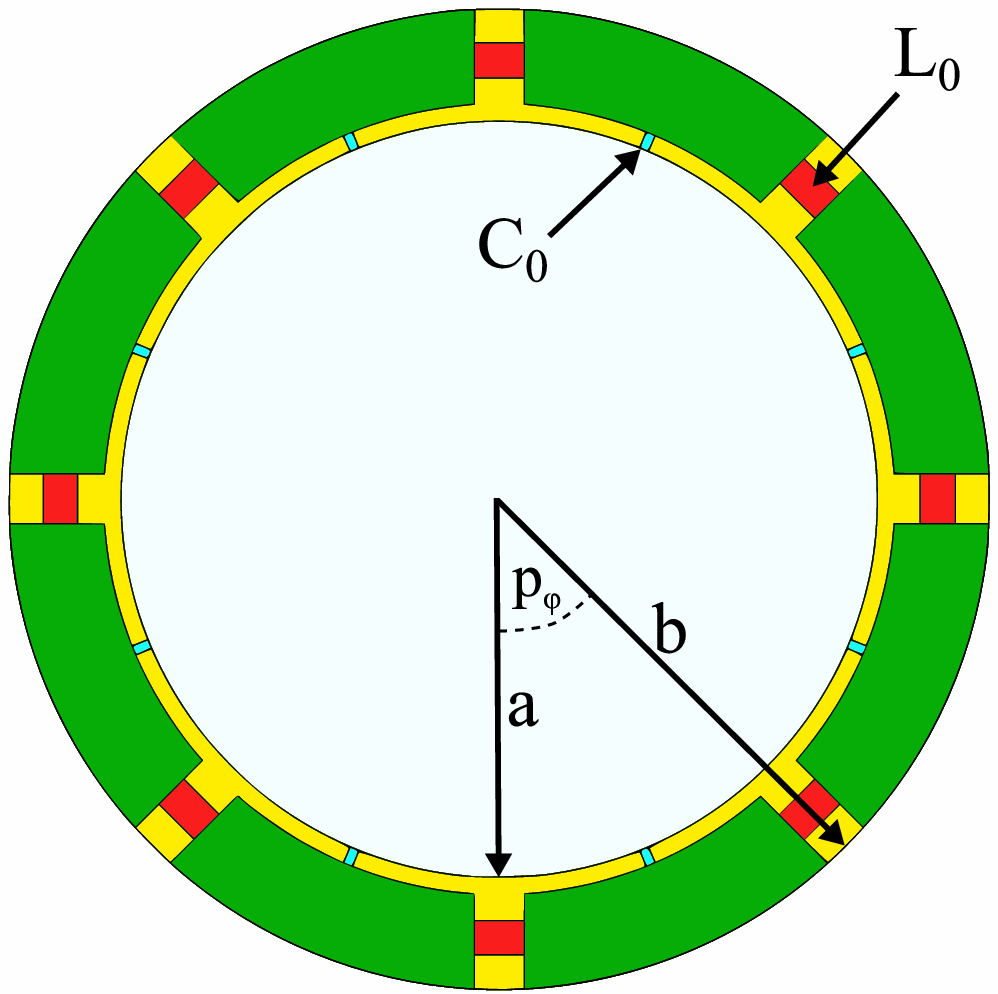}\label{fig3:subfig1}
}}
\subfigure[]{
\includegraphics[width=1.61in]{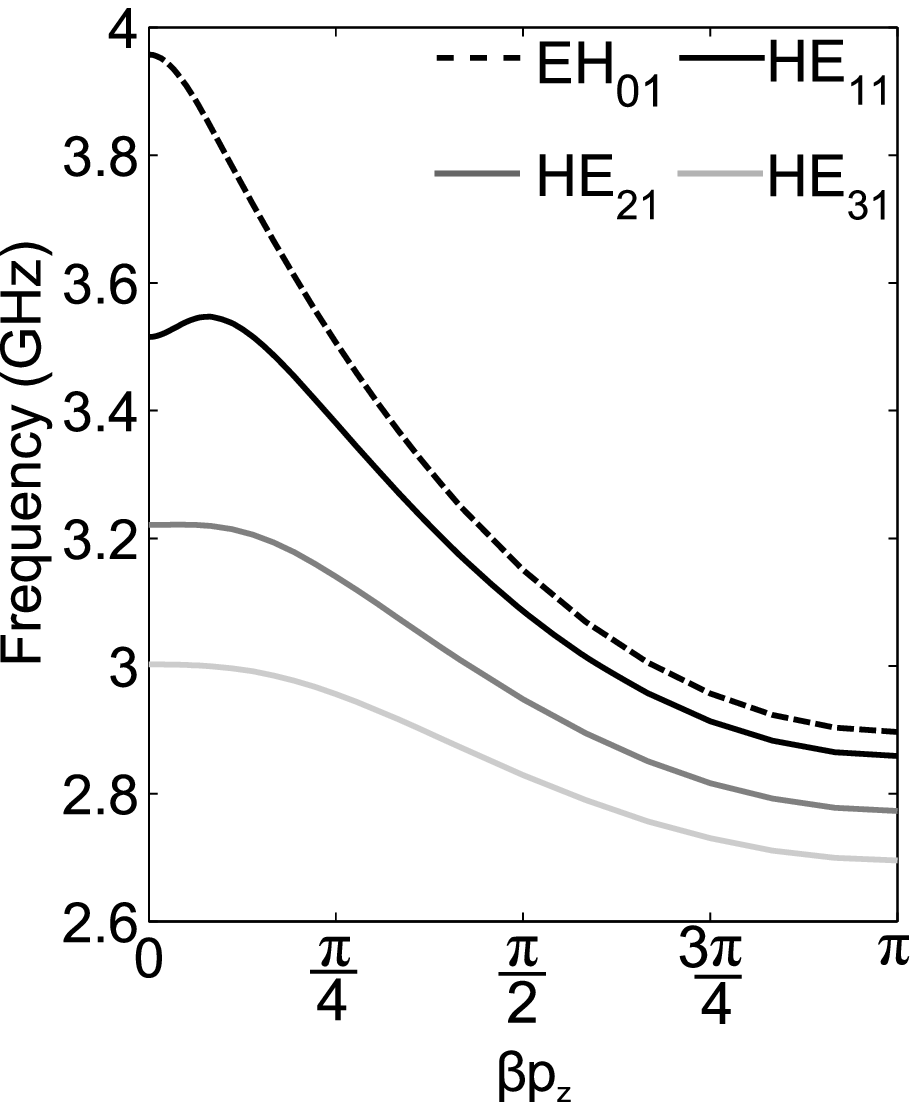}\label{fig3:subfig2}
}
\subfigure[]{
\includegraphics[width=3.25in]{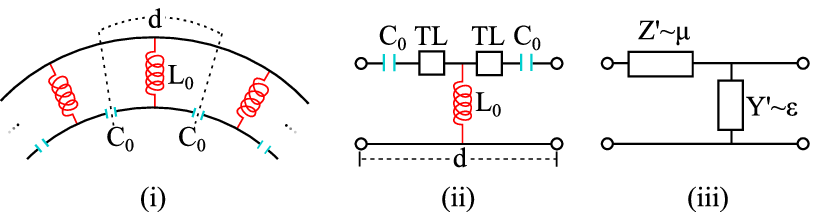}\label{fig3:subfig3}
}
\caption{\unskip (a) A representative printed-circuit implementation of an ENNZ metamaterial liner of thickness $t=b-a$ and (b) the dispersion of its frequency-reduced modes. (c) Evolution of a TL description of the printed-circuit implementation.}
\end{figure}

Figure~\ref{fig2} focuses on the frequency-reduced {\em EH}$_{01}$ and {\em HE}$_{11}$ modes. The normalized propagation ($\beta/k_0$) and attenuation ($\alpha/k_0$) constants are indicated by solid and dashed lines, respectively.
The frequency-reduced {\em EH}$_{01}$ mode (light grey curves) is described by a backward-wave band and, as mentioned earlier, a cutoff frequency $f_{01}=f_{ep}$. The anisotropic metamaterial-lined waveguide supports multiple highly dispersive frequency-reduced {\em HE}$_{n1}$ propagating bands for $n>0$. In addition to the waveguide-type {\em HE}$_{n1}$ modes, there also exist corresponding surface-type modes that are bound to the metamaterial-vacuum interface~\cite{novotny1994light}. These may be differentiated by their radial wavenumbers, $\gamma_{\rho 1}$. The waveguide-type modes exhibit a standing-wave distribution in the transverse plane due to an imaginary (i.e., propagating) $\gamma_{\rho1}$. The surface-type modes, on the other hand, decay evanescently into vacuum from the interface, and are characterized by real (i.e., attenuating) values of $\gamma_{\rho1}$. To retain clarity, Fig.~\ref{fig2} presents only the {\em HE}$_{11}$ solutions. In the observed range, the dispersion relation [Eqs.~(1)] reveals both the {\em HE}$_{11}$ mode's waveguide-type (black curves) and surface-type (dark grey circle curves) solutions, each exhibiting complex-mode regions for $f>f_{11}=3.70$GHz. Near $f_{11}$, both solutions have comparable $\alpha$ and $\beta$ values. This creates the potential for the backward-wave and complex bands to couple, and allows for regions of forward-wave propagation near cutoff. It is found that this effect may be mitigated by reducing the liner's thickness, increasing the degree of miniaturization, and choosing a lower-loss metamaterial. Below $f_{11}$, the two solutions rapidly become distinct, such that one can be classified as an evanescent band and another as a backward-wave propagating band. It is in this region that the metamaterial-lined waveguide shall be operated for transmission of the {\em HE}$_{11}$ mode.

\section{Realization of The ENNZ Liner}
\label{sec:real}
From the theoretical analysis in Sec.~\ref{sec:theory}, the frequency-reduced {\em HE}$_{11}$ mode's passband occurs when $\epsilon_{t2}$ achieves ENNZ values. This work realizes this ENNZ condition through the plasmonic-like interaction of arrays of thin wires~\cite{pendry1996}. Figure~\ref{fig3:subfig1} presents a printed-circuit-board (PCB) implementation of a single layer of the thin-wire metamaterial liner. Each layer consists of an orthogonal grid of radially and azimuthally directed thin copper traces loaded by discrete inductances $L_0$ and capacitive gaps modeled by discrete capacitances $C_0$, respectively. Layers are then stacked axially with a periodicity $p_z$. Although only a single $\rho$ period is used, the strong field confinement from the discrete inductors provides the desired miniaturization while allowing the liner to remain thin. The purpose of the discrete capacitors $C_0$ will be discussed in further detail shortly. Further details on the metamaterial design may be found in~\cite{pollockTAP2014}.
\subsection{Homogenization}
In general, this arrangement of thin wires can be modeled by a biaxial permittivity tensor whose transverse components ($\epsilon_{\rho2}$ and $\epsilon_{\phi2}$) are dispersive~\cite{demetriadou2008taming}. To model the impact of the transversely oriented current-carrying traces on the {\em HE}$_{11}$ mode's longitudinal magnetic field, $\mu_{z2}$ must also be assigned a dispersive response. Due to the absence of longitudinal wires and the small effect $\epsilon_{z2}$ has on the {\em HE}$_{11}$ mode's dispersion, we assign $\epsilon_{z2}=1$. Since it has further been established that $\epsilon_{\phi2}$ has a minimal impact on the cutoffs of {\em HE} modes, we retain the uniaxial assumption employed in Sec.~\ref{sec:theory} (i.e., $\epsilon_{t2}=\epsilon_{\rho2}=\epsilon_{\phi2}$) to aid in the homogenization. The PCB implementation is also assumed to have a nonmagnetic transverse response (i.e., $\mu_{\rho2}=\mu_{\phi2}=1$). We now describe a first-order homogenization procedure that may be used to specify the dispersive nature of $\epsilon_{t2}$ and $\mu_{z2}$, following which Eqs.~($1$) may be used to predict the modes' cutoffs.
It was shown in Sec.~\ref{sec:theory} that a frequency-reduced {\em EH}$_{01}$ mode cutoff frequency is introduced when $\epsilon_{t2}=0$. Observation of the fields in the practical metamaterial-lined waveguide reveals that, although they are predominantly {\em EH}$_{01}$-like in the vacuum region, the use of discrete inductive loading to realize a strongly near-zero permittivity and the introduction of azimuthally directed current-carrying copper traces results in a strong radial electric field and axial magnetic field in the liner region. These can be described as those of an azimuthally oriented TL mode supported between the copper trace and the waveguide wall. Here, this fact will be used to develop an equivalent-circuit model for the metamaterial liner based on TL metamaterial theory, from which its effective permittivity and permeability shall be extracted. The field polarizations of the TL mode suggest that these extracted parameters are intrinsically related to $\epsilon_{t2}$ and $\mu_{z2}$ of the liner.

Figure~\ref{fig3:subfig3} (i) shows a curved circuit description of multiple TL unit cells, in which $d=p_\phi(a+b)/2$ may be taken as an effective azimuthal periodicity. This is described by the equivalent-circuit model in Fig.~\ref{fig3:subfig3} (ii), which consists of a shunt inductor and series capacitor loading a host TL of periodicity $d$. The per-unit-length series impedance ($Z'$) and shunt admittance ($Y'$) of Fig.~\ref{fig3:subfig3} (iii) represent, respectively, the distributed inductance and capacitance of an effective TL describing the azimuthal TL mode. As mentioned, these distributed parameters can be related to the liner's effective-medium parameters, $\epsilon_{t2}$ and $\mu_{z2}$, whose dispersive (and assumed real) profile can be shown to take the following form:
\begin{subequations}\label{EQ7}
\begin{align}
\epsilon_{t2} = \epsilon_p ( 1 - \frac{{f_{ep}}^2}{f^2}), \;\;\;\;\;\mu_{z2} = \mu_p ( 1 - \frac{{f_{mp}}^2}{f^2}), \intertext{where}
f_{ep} = \frac{1}{2\pi}\sqrt{\frac{g}{\epsilon_p L_0 d}},\;\;\;\;\; f_{mp} = \frac{1}{2\pi}\sqrt{\frac{1}{g \mu_p C_0 d}}
\end{align}
\end{subequations}
In these expressions, $\epsilon_p$ and $\mu_p$ refer to the intrinsic material parameters of the host TL segments, $g$ is a geometrical parameter relating the TL mode's characteristic impedance to the wave impedance of the intrinsic medium, and $f_{ep}$ and $f_{mp}$ are the plasma frequencies at which $\epsilon_{t2}$ and $\mu_{z2}$ respectively achieve zero values.

To validate the proposed homogenization procedure, the representative design in Fig.~\ref{fig3:subfig1} with $0.3$mm- and $0.6$mm-wide radial and azimuthal copper traces, respectively, printed on a Rogers/Duroid $5880$ substrate, is assigned the following parameters: $b=15.0$mm, $a=12.7$mm, $L_0=8.4$nH, $C_0=0.01$pF, $p_\phi=45^\circ$, and $p_z=2.66$mm. To determine the parameters of the host TL, a flattened, unloaded, and similarly oriented copper trace of the same geometrical dimensions was simulated, revealing that $\epsilon_p=1.438$, $\mu_p=1$, and $g=0.7156$ at $f=4$GHz. For the specified loading values, $\epsilon_{t2}$ and $\mu_{z2}$ follow a Drude-like response with $f_{ep}=3.963$GHz and $f_{mp}=16.148$GHz, respectively. Whereas the strong shunt loading provides a frequency-reduced $f_{ep}$, the weak series capacitance $C_0$ produces a much greater $f_{mp}$, which seems to suggest that $\mu_{z2}$ assumes inordinately large negative values near $f_{ep}$. However, the Drude model for $\mu_{z2}$ is only accurate near $f_{mp}$. Near $f_{ep}$, $\mu_{z2}$ is better described by a single-pole Lorentz model with a resonance frequency $f_{m0}$. In this design, $C_0$ has been chosen small enough that $f_{m0}$ occurs well above $f_{ep}$, so that $\mu_{z2}$ here may be approximated as unity.

The homogenized liner and waveguide share the same physical dimensions as the PCB implementation ($b=15$mm and $a=12.7$mm). Using these values, the above homogenization procedure predicts frequency-reduced {\em EH}$_{01}$, {\em HE}$_{11}$, {\em HE}$_{21}$, and {\em HE}$_{31}$ modes with cutoff frequencies of $f_{01}=3.963$GHz, $f_{11}=3.512$GHz, $f_{21}=3.380$GHz, and $f_{31}=3.238$GHz, respectively. That $f_{01}=f_{ep}$ (i.e., where $\epsilon_{t2}=0$) makes physical sense from a TL perspective, since this implies an infinite wavelength condition in the TL mode, which is only satisfied for no azimuthal variation (i.e., $n=0$).

\begin{figure}[!t]
\centering
\subfigure[]{
\includegraphics[width=3in]{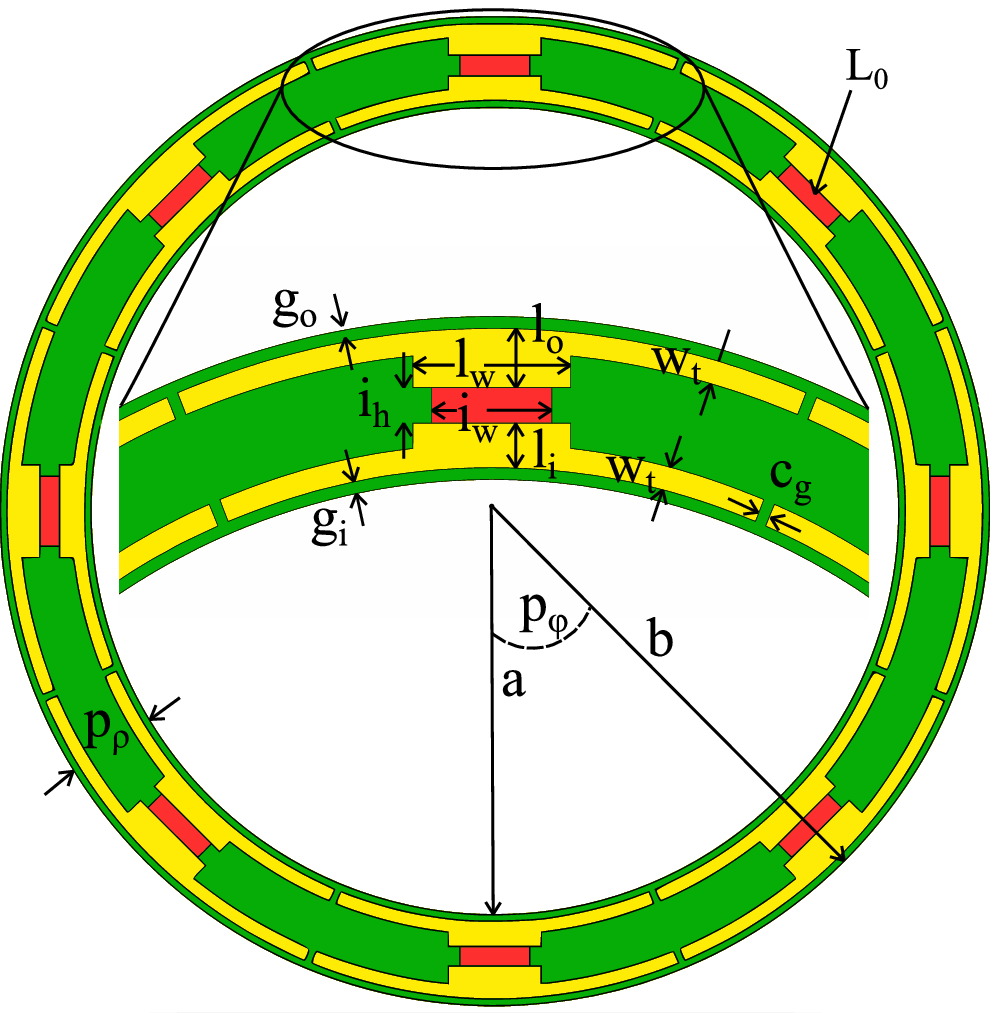}\label{fig4:subfig1}
}
\subfigure[]{
\includegraphics[width=3in]{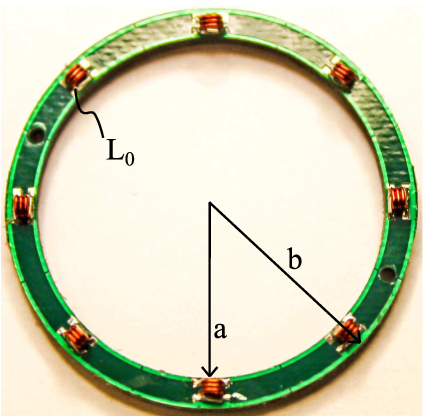}\label{fig4:subfig2}
}
\caption{\unskip (a) Simulation model and (b) fabricated prototype of the designed printed-circuit metamaterial based on inductively and capacitively loaded copper traces on a dielectric substrate.}
\end{figure}

Figure~\ref{fig3:subfig2} presents the dispersion of the axial phase incurred per unit cell ($\beta p_z$) obtained from HFSS's eigenmode simulator for the frequency-reduced {\em EH}$_{01}$ (black dashed curve), {\em HE}$_{11}$ (black solid curve), {\em HE}$_{21}$ (dark grey solid curve), and {\em HE}$_{31}$ (light grey solid curve) modes whose respective cutoff frequencies are $f_{01}=3.958$GHz, $f_{11}=3.515$GHz, $f_{21}=3.221$GHz, and $f_{31}=3.003$GHz, respectively. These are found to be within $1.2\%$, $0.08\%$, $4.8\%$, and $7.8\%$, respectively, of those predicted by the homogenization method. Interestingly, there remains a good agreement between the HFSS and homogenization results for the higher-azimuthal-order cutoffs, even though their associated large azimuthal phase variation significantly perturbs the assumption of homogeneity.

Although not shown, the homogenization results exhibit the same salient features as the HFSS data away from cutoff (backward-wave trends, cutoff frequencies decreasing with increasing azimuthal order, forward-backward coupling of the {\em HE}$_{11}$ mode). However, the latter appear less dispersive than the former. Here, it should be noted that the above homogenization approach only seeks to match the cutoff frequencies of the {\em EH}$_{01}$ and {\em HE}$_{11}$ modes, and not the full dispersion profile away from cutoff, for several reasons. Most significantly, the effective-medium approximation disregards the perturbations introduced by axial periodicity, which result in the coupling of forward and backward spatial harmonics at the band edges. Furthermore, the longitudinal electric fields become more pronounced away from cutoff, which deviates from the assumed TL-mode polarizations and also contributes to stronger spatial dispersion. Both effects break down the assumed Drude dispersion profile for $\epsilon_{t2}$~\cite{demetriadou2008taming}.

\begin{table}[!t]
\begin{center}\label{table:1}
\begin{tabular}{|l|cccccc|}
\multicolumn{7}{c}{TABLE I: \textsc{Liner Dimensions}} \\
\hline
(mm)&$b$&$a$&$c_g$&$w_t$&$g_0$&$g_i$\\
\hline
&15.0&12.59&0.10&0.30&0.10&0.15\\
\hline
(mm)&$l_0$&$l_i$&$i_h$&$l_w$&$i_w$&\\
\hline
&1.00&0.65&0.51&2.60&1.83&\\
\lasthline
\end{tabular}
\end{center}
\end{table}
\subsection{Fabricated Metamaterial Liner}
Figure~\ref{fig4:subfig1} presents a version of the printed-circuit metamaterial that is amenable to standard PCB fabrication. A small radial gap of width $g_0$ and an outer $\phi$-directed trace of width $w_t$ are introduced. The outer trace replaces the waveguide wall's function in establishing a continuous azimuthal current path between adjacent $\rho$-directed inductors. This eases the fabrication challenge of soldering to the waveguide wall, which was required in the previous design, and enables the ENNZ liner to be modular. In this design, Coilcraft ultra-high-Q 0806SQ-6N0 inductors were chosen. Their inductance of $L_0=8.28$nH and  quality factor $Q=100$ at $f=3.70$GHz were extrapolated outside the frequency range presented in their data sheets~\cite{coilcraft}. Due to the larger concentration of fields around each inductor, the value of $Q$ is a pivotal factor in defining the {\em HE}$_{11}$ mode's propagation loss. The traces are printed on a Rogers/Duroid $5880$ substrate of $61$mil thickness which, with the chosen inductor's height, gives $p_z=3.017$mm. The remaining design dimensions shown in Fig.~\ref{fig4:subfig1} are listed in Table I, and were chosen to provide a reasonably close match to the cutoffs of the previous design shown in Fig.~\ref{fig3:subfig1}.

\begin{figure}[!t]
\centering
\includegraphics[width=3.1in]{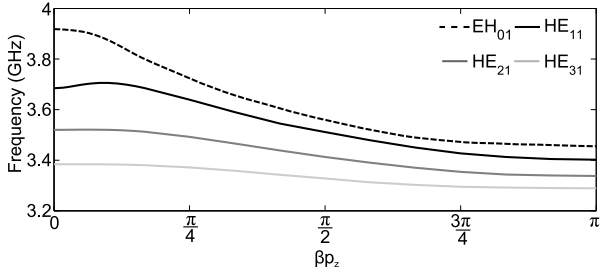}
\caption{\unskip  Dispersion of the frequency-reduced {\em EH}$_{01}$, {\em HE}$_{11}$, {\em HE}$_{21}$, and {\em HE}$_{31}$ modes as obtained from full-wave eigenmode simulations.\label{fig5}}
\end{figure}

A photo of the fabricated PCB metamaterial is shown in Fig.~\ref{fig4:subfig2}. Full-wave eigenmode simulations of this structure reveal that it supports frequency-reduced {\em EH}$_{01}$, {\em HE}$_{11}$, {\em HE}$_{21}$, and {\em HE}$_{31}$ modes, as shown in Fig.~\ref{fig5}, with cutoff frequencies of $f_{01}=3.918$GHz, $f_{11}=3.685$GHz, $f_{21}=3.520$GHz, and $f_{31}=3.384$GHz, respectively. However, these simulated data do not include the impact of fabrication tolerances imposed through milling, etching, and component placement.

\begin{figure}[!t]
\centering
\subfigure[]{
\includegraphics[width=1.625in]{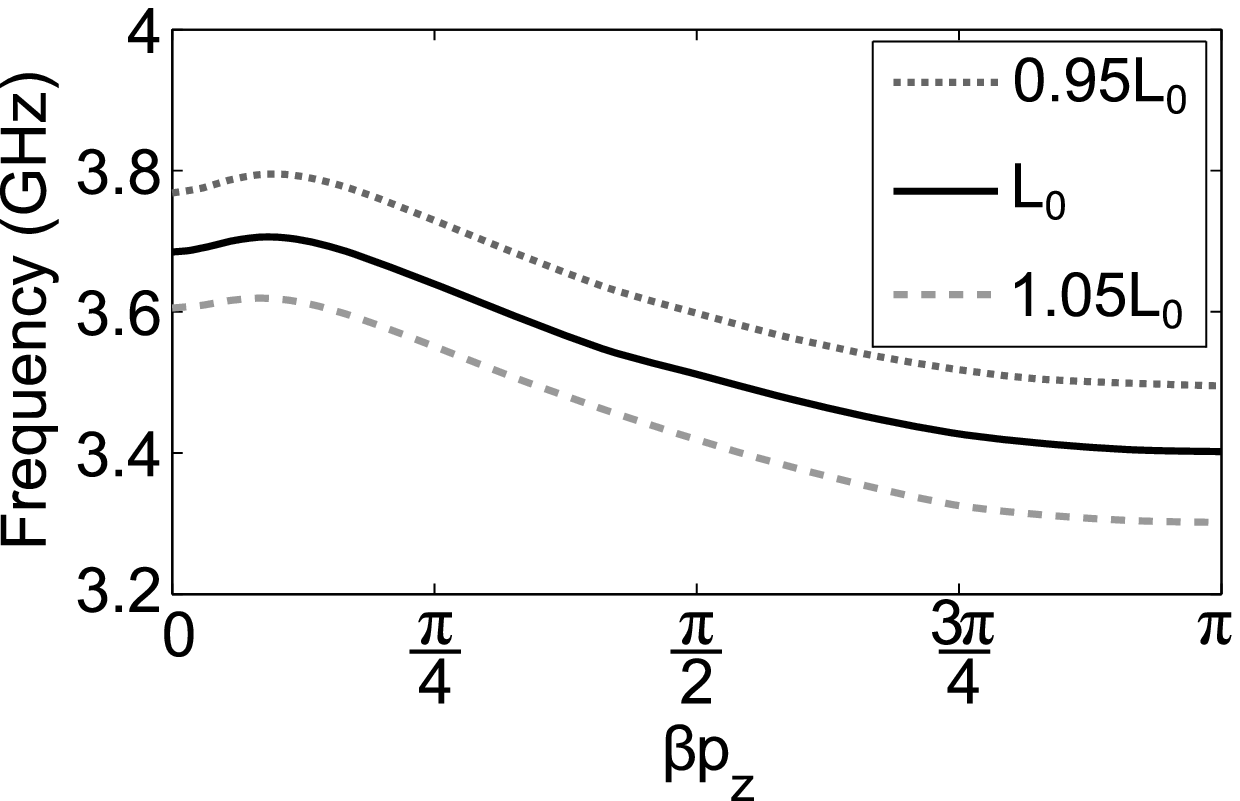}
\label{fig6:subfig1}
}
\subfigure[]{
\includegraphics[width=1.625in]{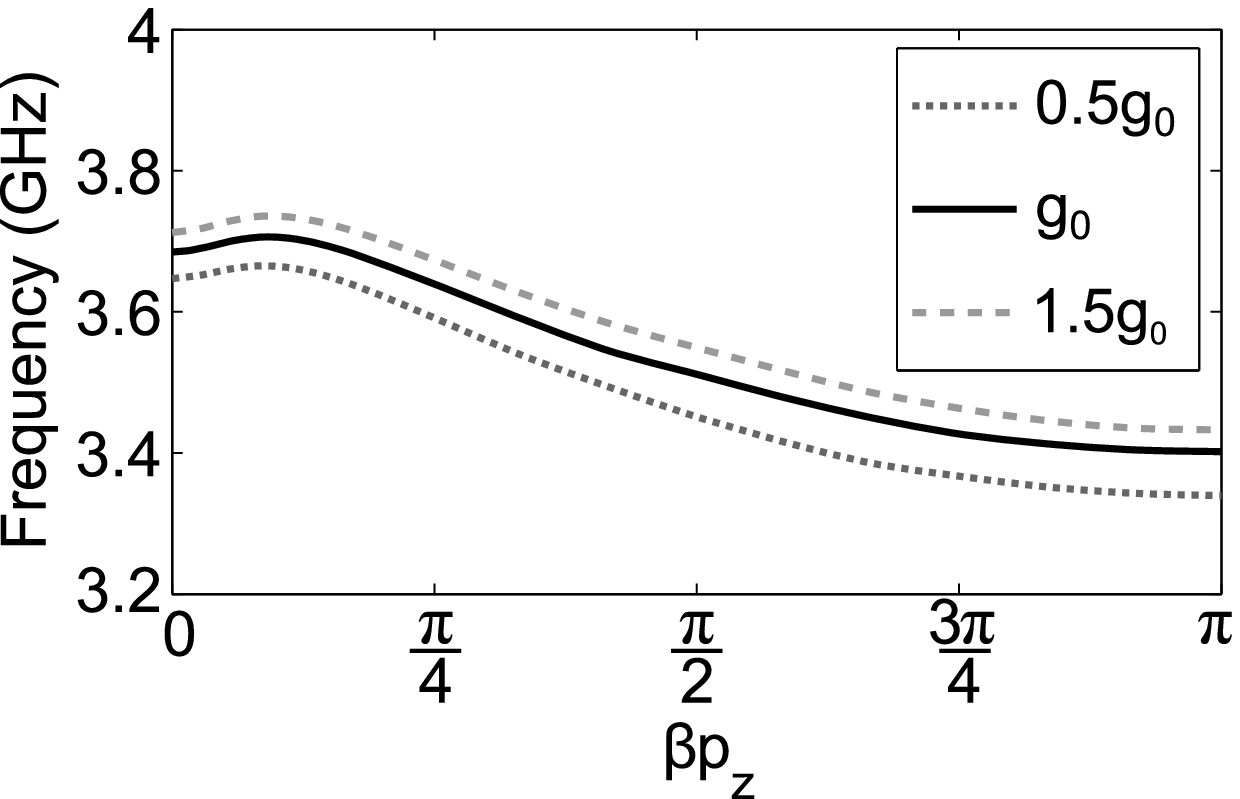}
\label{fig6:subfig2}
}
\subfigure[]{
\includegraphics[width=1.625in]{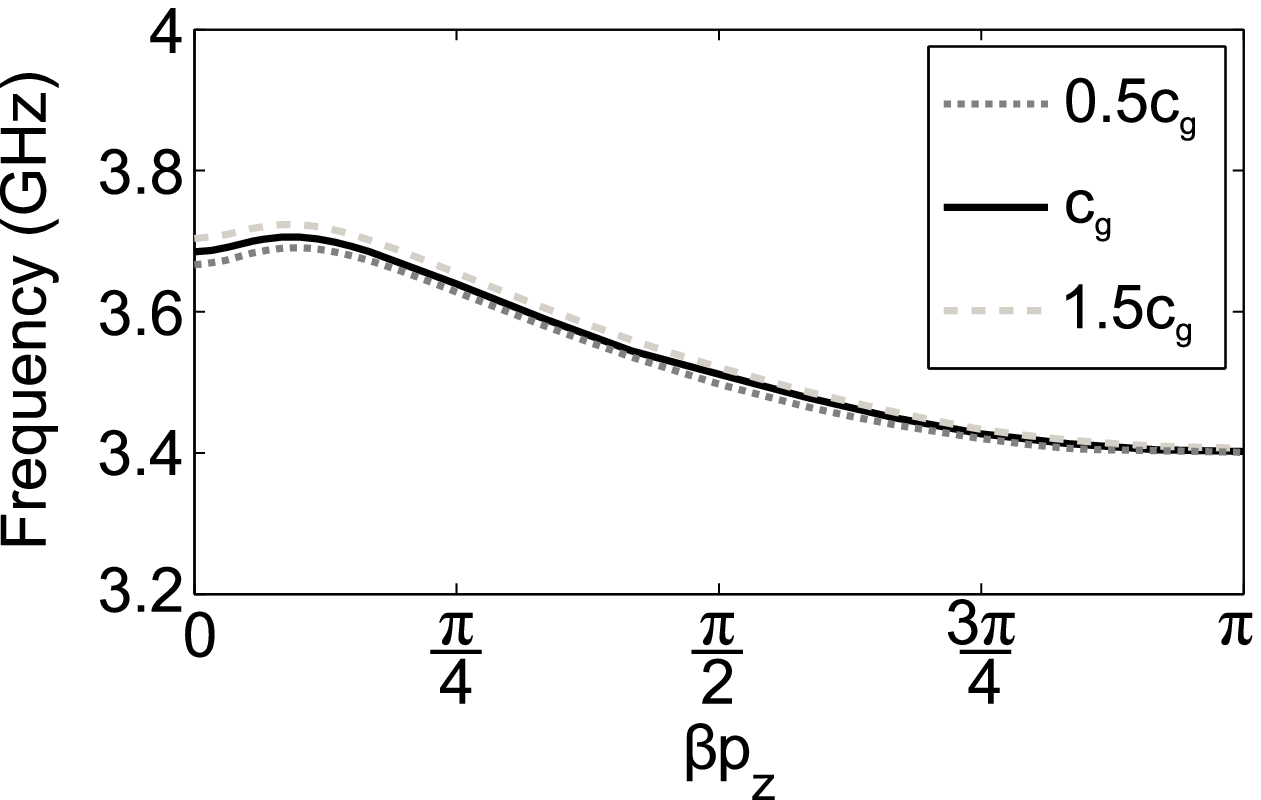}
\label{fig6:subfig3}
}
\subfigure[]{
\includegraphics[width=1.625in]{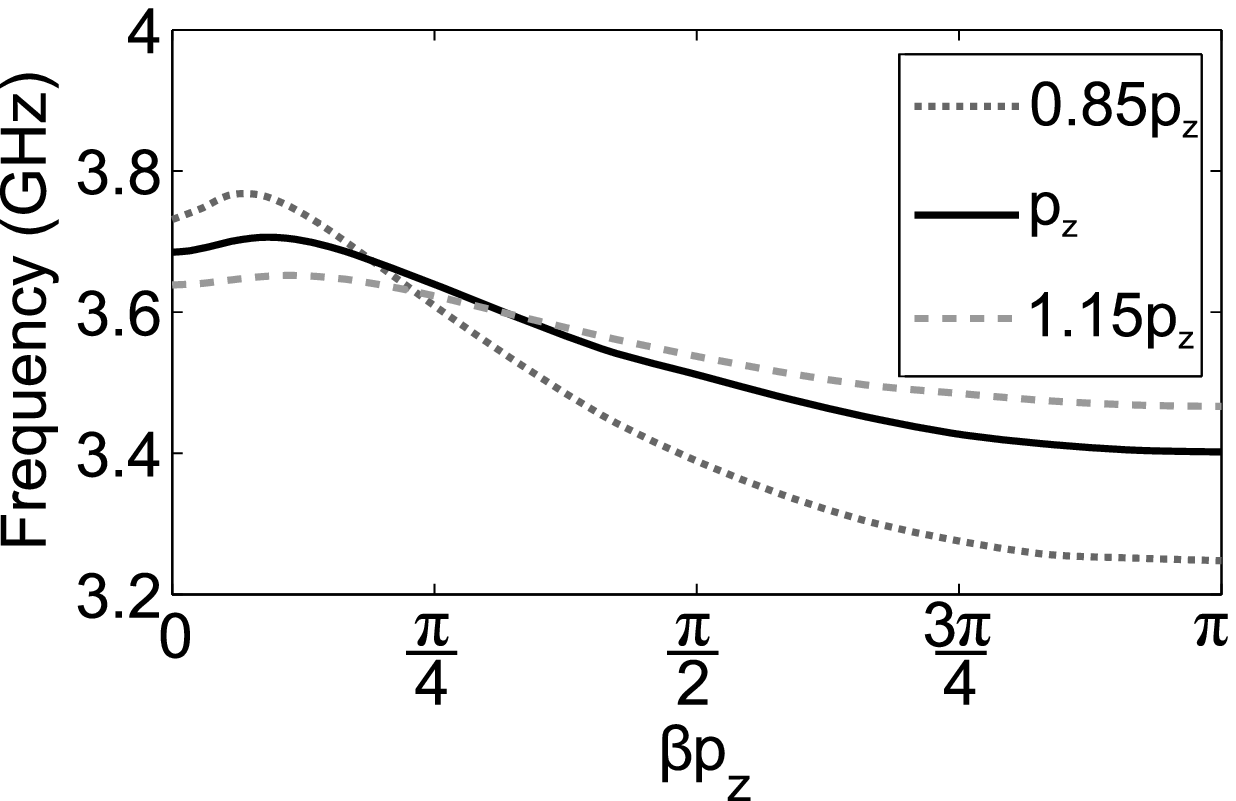}
\label{fig6:subfig4}
}
\caption{\unskip Variation of the dispersion of the frequency-reduced {\em HE}$_{11}$ mode with (a) $L_0$, (b) $g_0$, (c) $c_g$, and (d) $p_z$.}
\end{figure}

Figures~\ref{fig6:subfig1}-\ref{fig6:subfig4} present the dispersion curves of the frequency-reduced {\em HE}$_{11}$ mode for varying $L_0$, $p_z$, $c_g$, and $g_0$ values, respectively. Since the loading inductance, $L_0$, primarily controls the location of the frequency-reduced mode cutoffs, it is instructive to look at the impact of their listed tolerances of up to $5\%$ ($0.414$nH). Furthermore, precise measurements of each the fabricated layers' $p_z$ reveal tolerances of up to $15\%$ ($0.45$mm) per layer. Last, the metamaterial's extremely fine features push the limits of the milling and etching process and this leads to tolerances of up to $50\%$ ($50\mu$m) in $c_g$ and $g_0$. Each figure shares the nominal case (solid black curve) whose dimensions were presented in Table I. While all the observed parameters shift the band up or down, $L_0$ commands the greatest change in the cutoff frequency. The cutoff is only moderately impacted by $g_0$ and $p_z$ and only minutely by $c_g$. The latter is to be expected from the equivalent-circuit model, which predicts that small changes in the capacitance $C_0$ only impact $\mu_{z2}$ near its plasma frequency (i.e., for $f>>f_{ep}$). The dispersion profile's shape is dependent on $p_z$ and is more dispersive when the unit cells are stacked closer together. It is worth noting that, despite the tolerance levels of each parameter, the {\em HE}$_{11}$ mode's cutoff frequency always remains within $3\%$ of its designed value of $f_{11}=3.685$GHz.

\section{Numerical and Experimental Verification}
\label{sec:full}
\begin{figure}[!t]
\centering
\subfigure[]{
\includegraphics[width=3.1in]{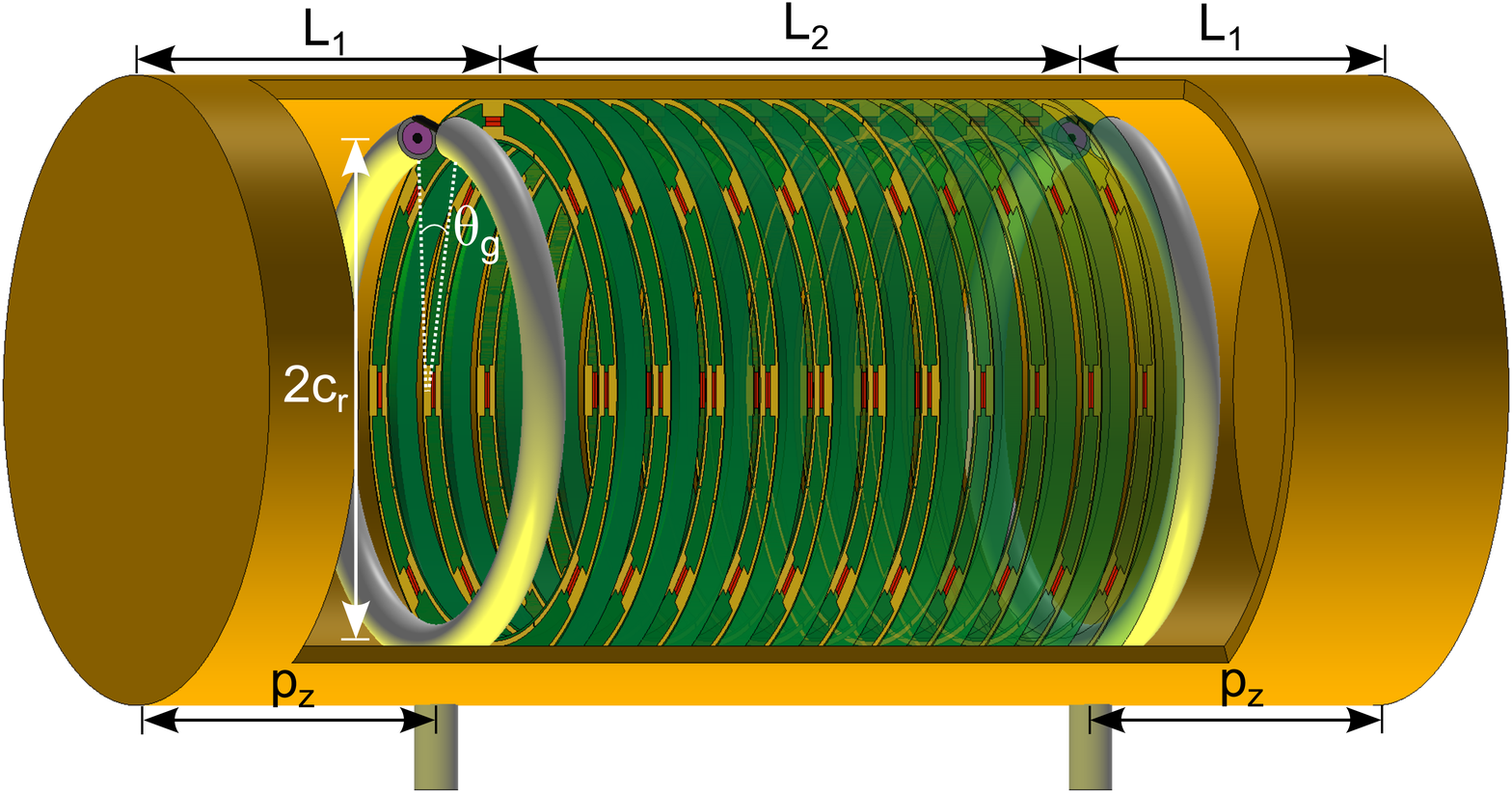}\label{fig7:subfig1}
}
\subfigure[]{
\includegraphics[width=3.2in]{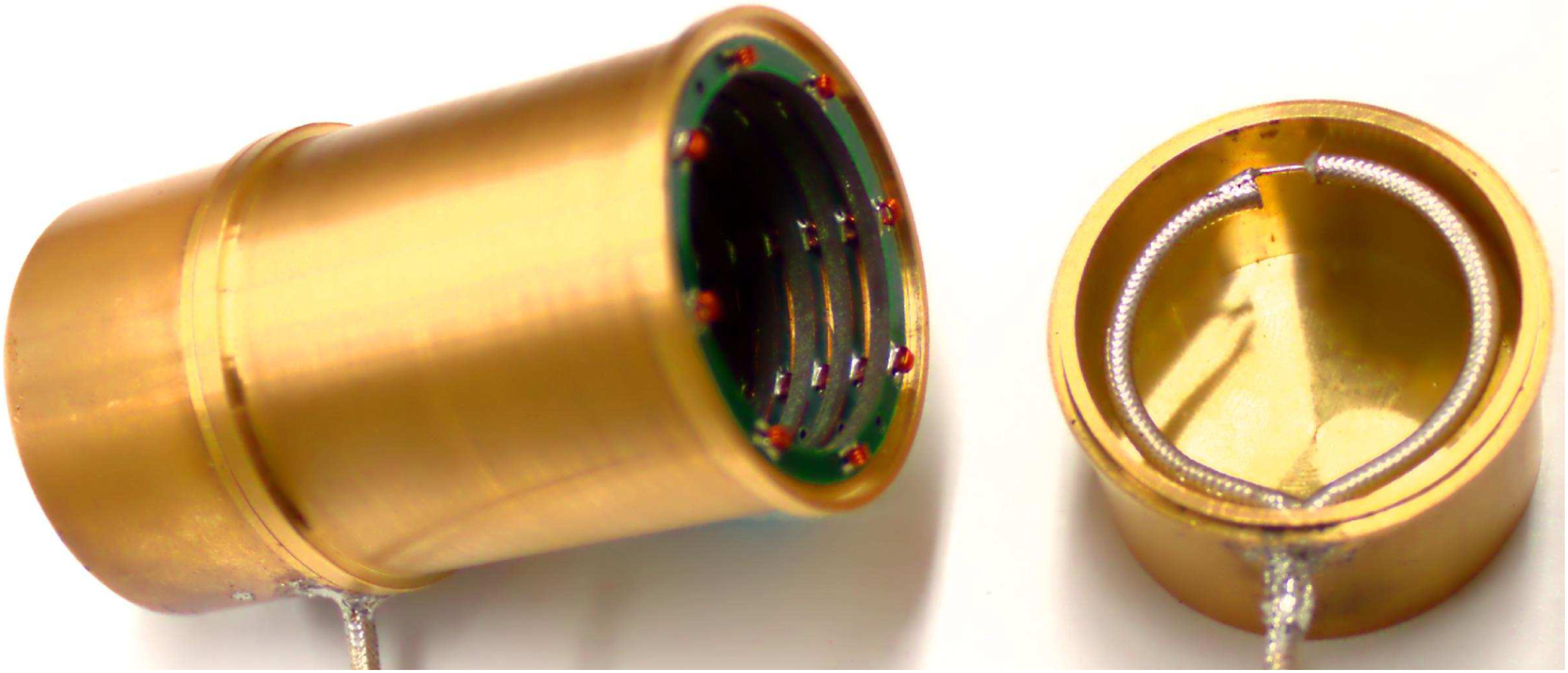}\label{fig7:subfig2}
}
\caption{\unskip (a) Full-wave simulation model and (b) fabricated experimental prototype of the metamaterial-lined waveguide showing one of two shielded-loop-antenna sources.}
\end{figure}

Figures~\ref{fig7:subfig1} and~\ref{fig7:subfig2} present the full-wave-simulation and experimental setup, respectively, which consist of two vacuum-filled excitation waveguides of radius $b=15$mm and a {\em TE}$_{11}$ cutoff frequency of $5.857$GHz that are smoothly connected to an intermediate metamaterial-lined waveguide. In simulation studies describing the liner as isotropic and homogeneous, the transmission through a below-cutoff waveguide was significantly improved~\cite{pollockmttt2013}. In this work, the PCB metamaterial presented in Figs.~\ref{fig4:subfig1} and~\ref{fig4:subfig2} is arranged into a stack of $11$ layers along the intermediate waveguide's length of $L_2=32$mm. To efficiently excite the {\em HE}$_{11}$ mode without strongly exciting the {\em EH}$_{01}$ and other higher-order frequency-reduced modes, two balanced shielded loops~\cite{whiteside1964loop} are embedded within the closed evanescent waveguide sections of length $L_1=17.75$mm at a distance $p_z=15.0$mm from a PEC back wall. Details of the design for each shielded loop (radius $c_r=13.5$mm  and gap $\theta_g=20^\circ$) can be found in~\cite{pollockTAP2014}.
\begin{figure}[!t]
\centering
\subfigure[]{
\includegraphics[width=3.25in]{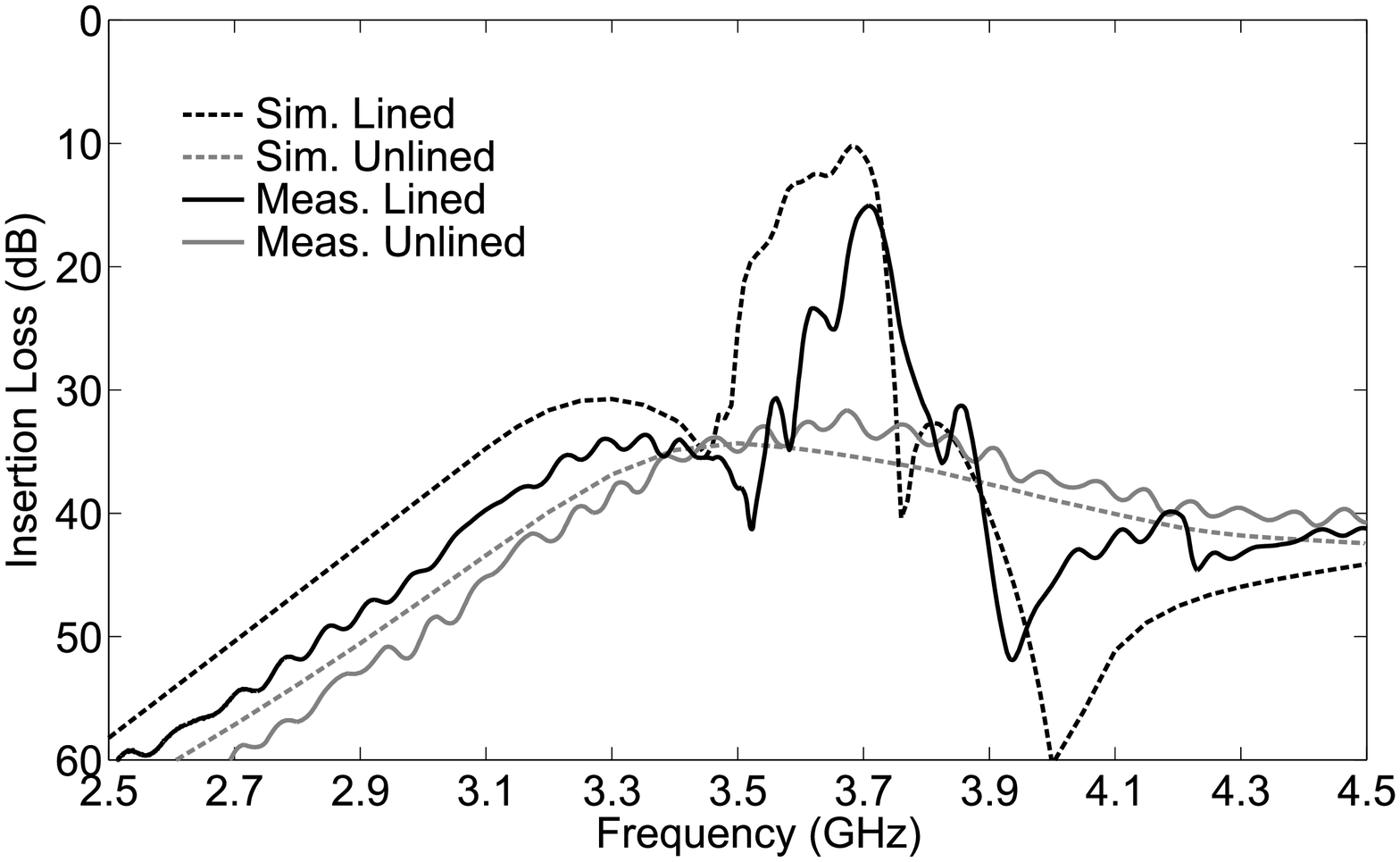}\label{fig8:subfig1}
}
\subfigure[]{
\includegraphics[width=3.25in]{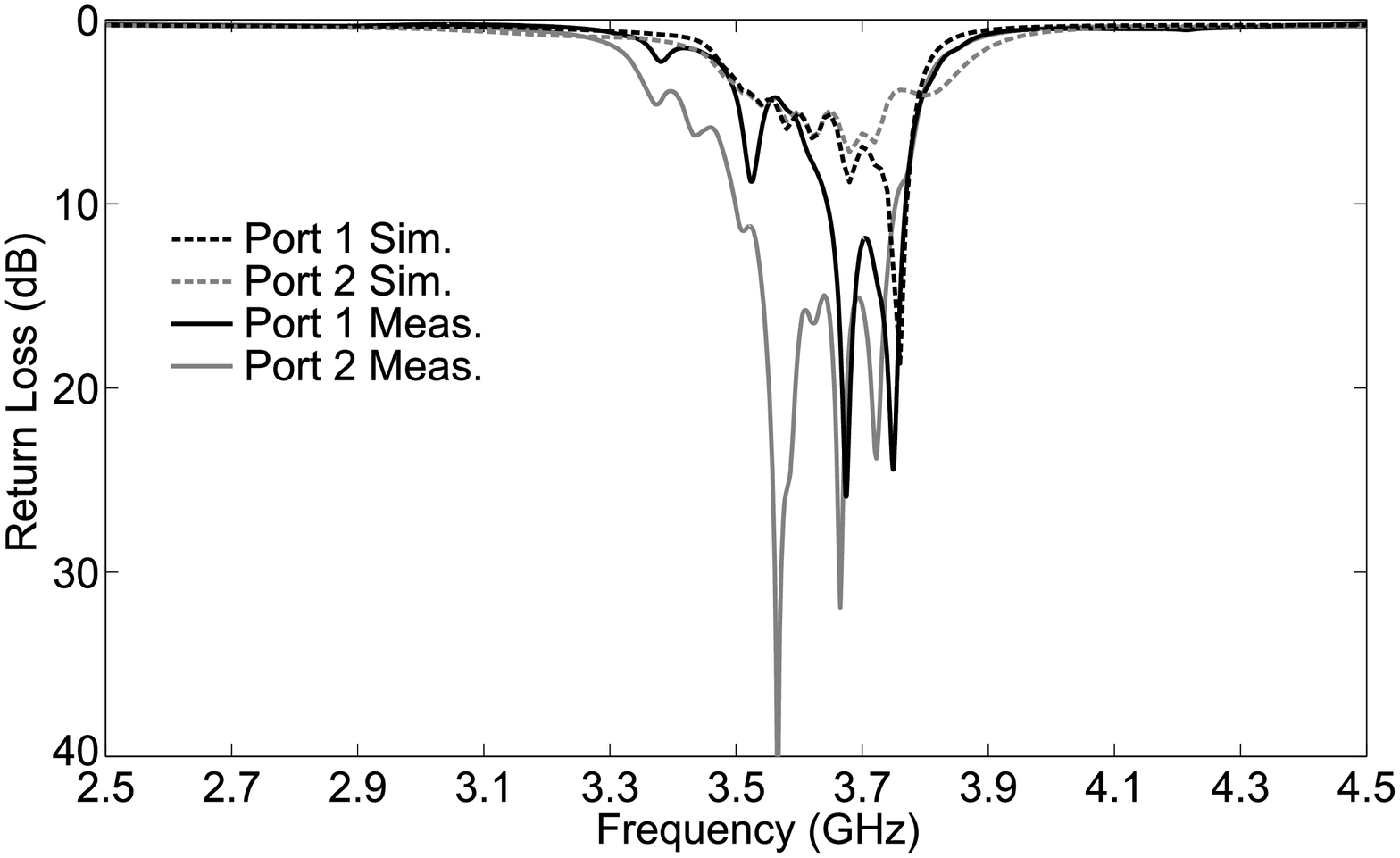}\label{fig8:subfig2}
}
\caption{\unskip (a) Insertion loss and (b) return loss as obtained from simulation and measurements for the transmission setup depicted in Figs.~\ref{fig7:subfig1} and~\ref{fig7:subfig2}.}
\end{figure}
\subsection{Transmission Simulations}
\begin{figure}[!t]
\centering
\includegraphics[width=3in]{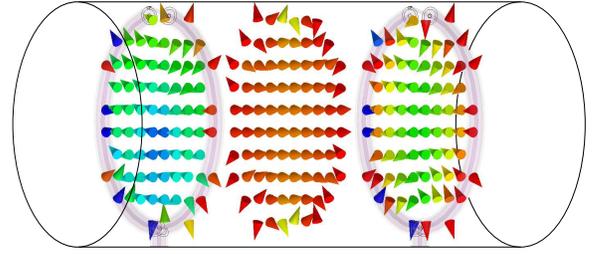}
\caption{Simulated complex electric-field vectors at different planes at $f=3.720$GHz showing excitation and detection of {\em TE}$_{11}$ modes coupled through an {\em HE}$_{11}$ mode supported by the metamaterial-lined waveguide.\label{fig8:subfig3}}
\end{figure}

Without the metamaterial liner, the unloaded vacuum-filled intermediate waveguide is within its natural evanescent region for frequencies below $f=5.857$GHz. The dashed grey curve in Fig.~\ref{fig8:subfig1} presents the insertion loss obtained from simulations for this case, and verifies that the intermediate waveguide under cutoff strongly attenuates the {\em TE}$_{11}$ mode. The insertion loss achieves a minimum of $34$dB at $f=3.500$GHz. Now, according to Fig.~\ref{fig5}, introducing the PCB metamaterial layers should enable {\em HE}$_{11}$ propagation below $f_{11}=3.685$GHz by way of a frequency-reduced backward-wave band. HFSS full-wave simulations show that the insertion loss (black dashed curve in Fig.~\ref{fig8:subfig1}) has a passband whose upper-band edge is roughly situated near $f_{11}$ and which demonstrates improvements in the transmission by up to $25.4$dB (at $f=3.680$GHz). The dashed curves in Fig.~\ref{fig8:subfig2} show the return loss of the simulated metamaterial-lined waveguide. It should be noted that the metamaterial layers are not symmetrical, since the shielded loops at ports $1$ and $2$ respectively face the inductor or the substrate dielectric. This asymmetry translates to unequal return loss profiles at Port $1$ (grey dashed curves) and Port $2$ (black dashed curves). Nevertheless, the return loss of both ports is described by several resonant peaks, and achieves maximum values of $18.7$dB at $f=3.760$GHz (Port $1$) and $7.2$dB at $f=3.680$GHz (Port $2$). Figure~\ref{fig8:subfig3} shows the simulated electric-field vectors at $f=3.720$GHz located at the following cross sections:  the excitation shielded loop, the center of the metamaterial-lined region, and the receiving shielded loop. The field patterns confirm that the shielded loop strongly excites and receives the {\em TE}$_{11}$ mode, which is coupled from the input to output waveguide sections through the {\em HE}$_{11}$ mode in the metamaterial-lined waveguide.

\begin{figure}[!t]
\centering
\subfigure[]{
\includegraphics[width=1.6in]{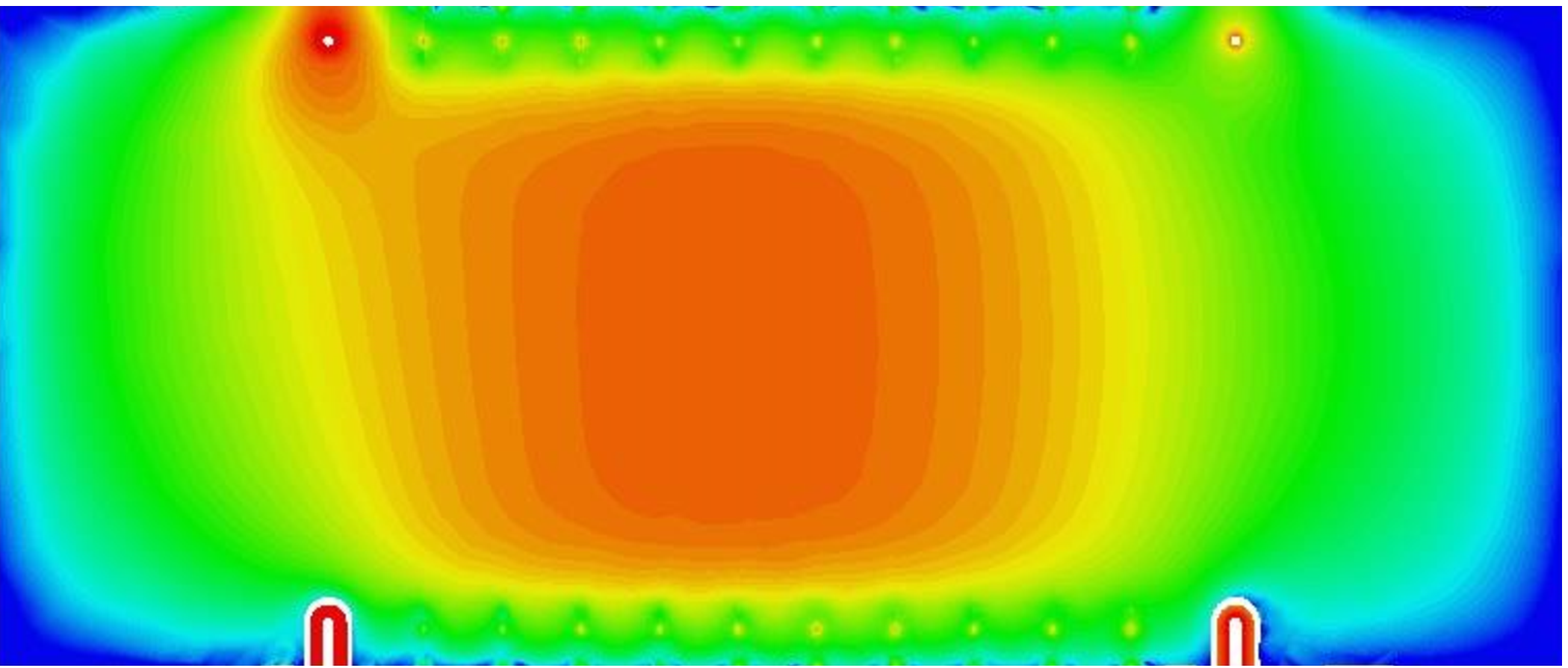}\label{fig9:subfig1}
}
\subfigure[]{
\includegraphics[width=1.6in]{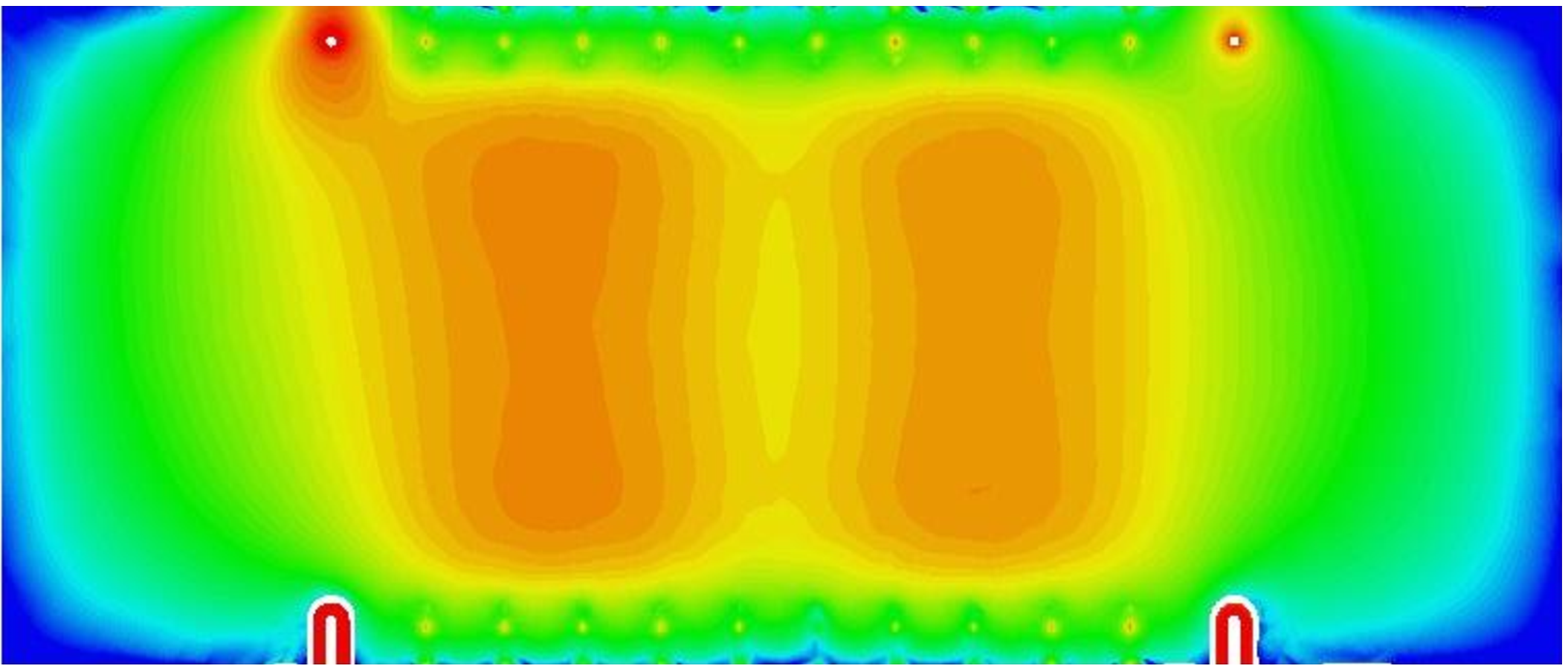}\label{fig9:subfig2}
}
\subfigure[]{
\includegraphics[width=1.6in]{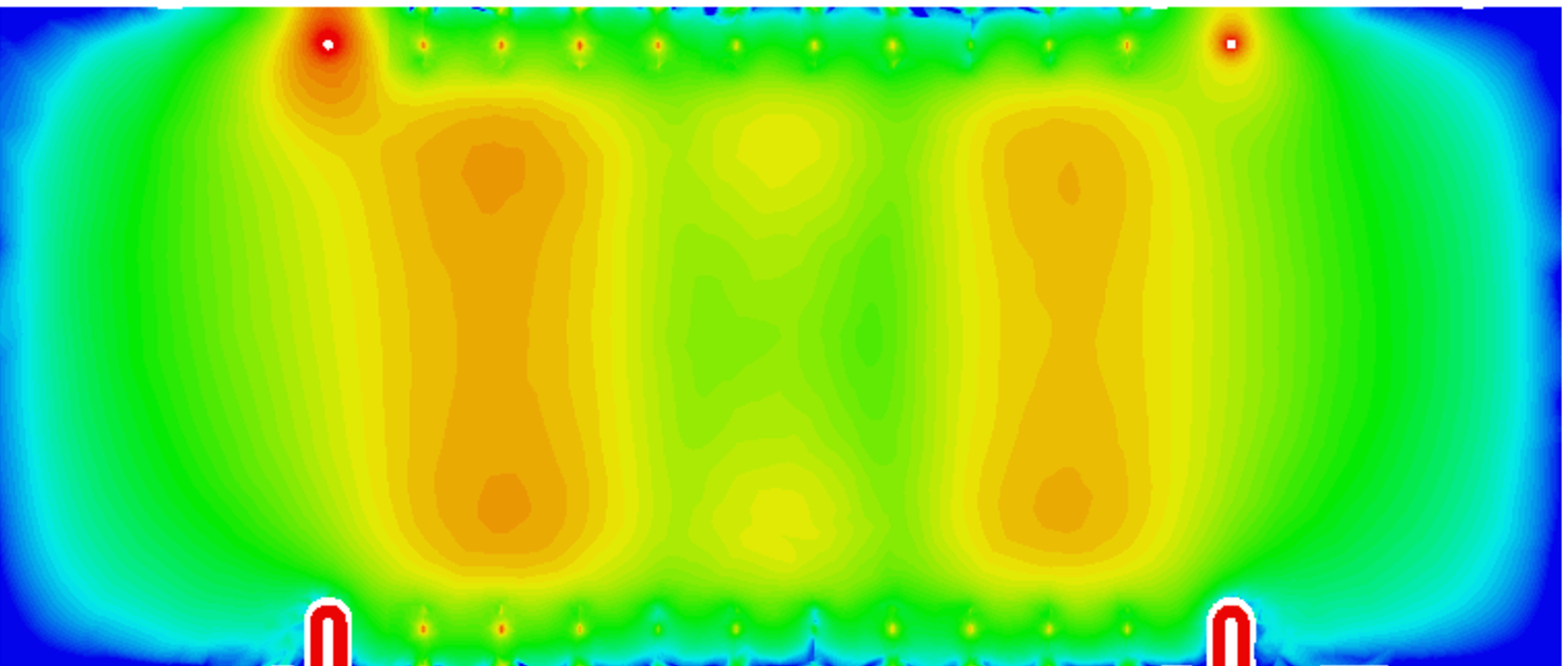}\label{fig9:subfig3}
}
\subfigure[]{
\includegraphics[width=1.6in]{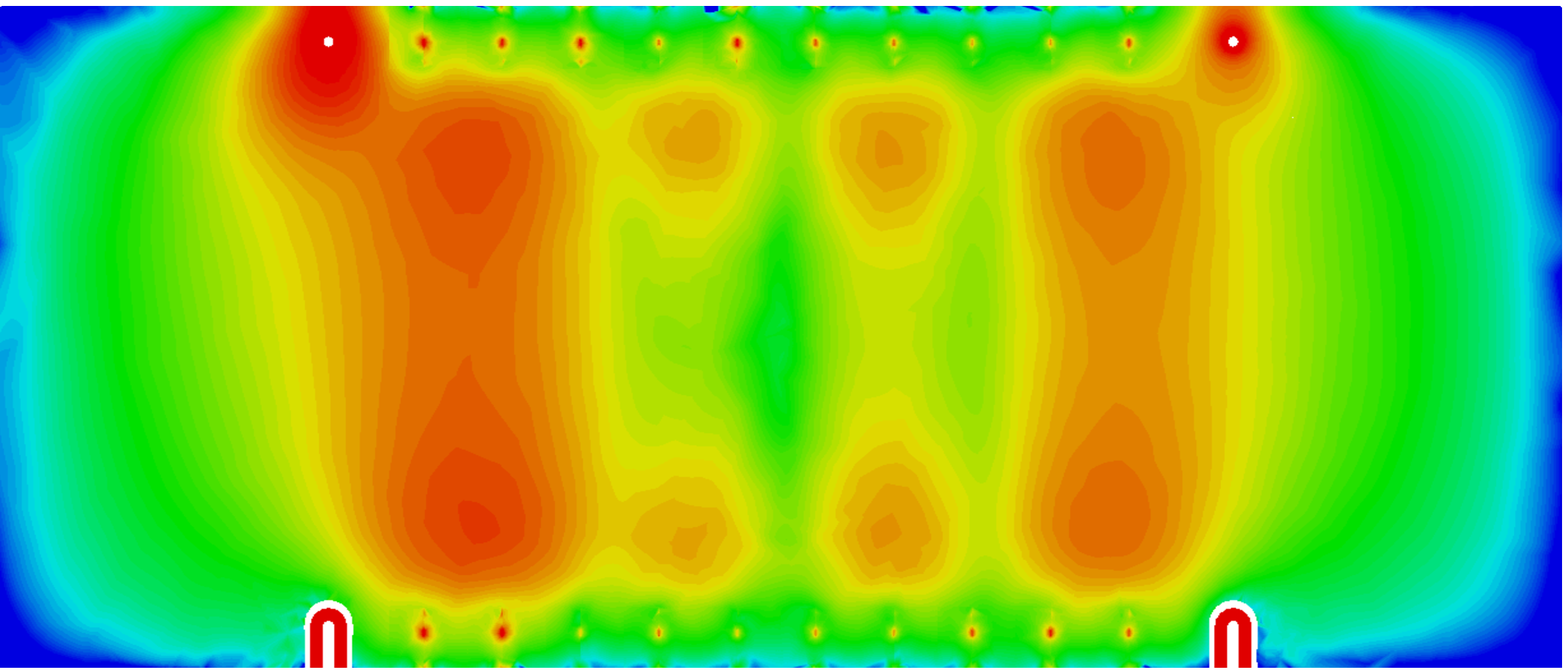}\label{fig9:subfig4}
}
\caption{\unskip Simulated complex electric-field magnitudes shown in the metamaterial-lined waveguide at the following transmission peaks in the frequency-reduced backward-wave passband: (a) $f=3.720$GHz, (b) $f=3.680$GHz, (c) $f=3.620$GHz, and (d) $f=3.580$GHz.}
\end{figure}

Figures~\ref{fig9:subfig1}--\ref{fig9:subfig4} present the simulated complex electric-field magnitudes for the transmission setup of Fig.~\ref{fig7:subfig1} at the first four transmission peaks located at $f=3.720$GHz, $f=3.680$GHz, $f=3.620$GHz, and $f=3.580$GHz, respectively. It is instructive to compare these field data to those presented in~\cite{pollockmttt2013} for the isotropic, homogeneous ENNZ liner. They are similar in that each successive peak corresponds to a Fabry-P\'erot-type resonance of the {\em HE}$_{11}$ mode in which an integer number of half-wavelengths is supported by the lined waveguide section over its length. Indeed, the resonant nature of the transmission is readily observed in Figs.~\ref{fig9:subfig1}--\ref{fig9:subfig4}. However, they differ in that the PCB metamaterial exhibits an electric field that is more evenly distributed across the waveguide's cross section and an overall reduced bandwidth of the passband. These differences are attributed to the periodic nature of the PCB implementation, which relaxes the strict boundary conditions enforced on the electric fields at the metamaterial-vacuum interfaces. Furthermore, as noted in Sec.~\ref{sec:real} A, the finite periodicity results in a flattening of the {\em HE}$_{11}$ dispersion, which therefore limits the lower edge of the propagating band. Nevertheless, across the PCB metamaterial liner's backward-wave band, the level of transmission is generally better than that of the homogeneous ENNZ liners. This improvement can be attributed to the decreased insertion loss over the shorter waveguide length and an improvement in matching owed to the use of the shielded-loop excitation.
\subsection{Transmission Measurements}
A Keysight PNA-X N5244A Network Analyzer was used to measure the response of the experimental prototype shown in Fig.~\ref{fig7:subfig2}. For the unloaded vacuum-filled waveguide, the measured insertion loss (solid grey curve in Fig.~\ref{fig8:subfig1}) shows a general agreement with simulations, with a minimum of $32$dB at $f=3.676$GHz. This $4.8\%$ upshift in its optimal transmission frequency with respect to simulations is attributed to the fabricated shielded loop's slightly smaller radius ($c_r$) and tolerances in realizing the gap ($\theta_g$) by hand. The modular nature of the multi-layered PCB metamaterial allows it to now be easily inserted into the intermediate waveguide. Pins are used to ensure the position and alignment is kept consistent with the simulation model, and are removed before measurements. The measured insertion loss (solid black curve) shows a very good agreement with simulations in which both data exhibit the same salient features. These include an anti-resonance at frequencies above an appreciable passband, below which there is a moderate roll-off in the insertion loss. This passband achieves an insertion loss as low as $15.0$dB at $f=3.710$GHz, which is only a $4.8$dB decrease and $0.8\%$ frequency upshift with respect to the optimal transmission in simulations. This represents a $19$dB enhancement in transmission over the unloaded waveguide. It should be noted that, at this frequency, the metamaterial-lined waveguide's cross-sectional area is $60\%$ smaller than that of a vacuum-filled circular waveguide operating at cutoff. This is a representative example of the even more extreme miniaturizations that may be achieved with ENNZ metamaterial liners. The measured return loss in Fig.~\ref{fig8:subfig2} for Port $1$ (solid black curve) and Port $2$ (solid grey curve) generally agree with those from simulations in the passband, but achieve better performance. Port $1$ and Port $2$ achieve a maximum return loss of $18.7$dB at $f=3.760$GHz and $7.2$dB at $f=3.680$GHz, respectively, and $10$dB bandwidths of $3.3\%$ and $7.0\%$, respectively.

The general agreement of the passband region between simulations and measurements occurs despite the fabrication tolerances imposed on the shielded loop and PCB metamaterial. This is to be expected from the discussion in Sec.~\ref{sec:real}, which highlights that, for the expected tolerances on $L_0$, $g_0$, $c_g$, and $p_z$, the {\em HE}$_{11}$ mode's dispersion remains roughly unchanged. However, the cumulative effect of these imperfections imposed on the multilayered stack will increase insertion loss due to decreased coupling between layers, particularly at lower frequencies in the backward-wave band where the unit cells lose their homogeneity. Furthermore, the inductor's loaded quality factor in the fabricated unit cell may have been lower than approximated in simulations, which increases the insertion loss in the passband while preserving its general profile. Although more accurate simulations would have lowered the discrepancy, the computational intensity to achieve this was not feasible. It is expected that the observed transmission may be improved by reducing the loss in the liner region using either higher-Q inductors or increasing liner thickness to minimize field confinement. Of course, as the liner becomes thicker (i.e., $a\rightarrow0$), the waveguide becomes homogeneously ENNZ-filled and transmission is expected to become independent of its geometry\cite{enghati_tunnel}.

\section{Conclusion}
The propagation characteristics of a miniaturized circular waveguide containing an anisotropic metamaterial liner have been analyzed theoretically, numerically, and experimentally. A full anisotropic treatment of the liner reveals that a spectrum of frequency-reduced modes are introduced well below its natural cutoff frequency, where the liner's transverse permittivity assumes negative and near-zero values. It was shown how the cutoff frequency of the {\em HE}$_{n1}$ modes may be designed using the dimensions and permittivities of the liner and waveguide, while the cutoff frequency of the {\em EH}$_{01}$ mode occurs at the liner's plasma frequency. A practical PCB-based metamaterial based on inductively loaded metallic traces is shown to yield dispersion and miniaturization properties that are consistent with those observed for homogeneous, anisotropic metamaterial liners. A first-order homogenization procedure used to model the PCB implementation employs an effective anisotropic permittivity and is shown to accurately match the cutoff frequencies of the lowest order frequency-reduced modes. Full-wave simulations and experimental results were presented for a waveguide lined with $11$ metamaterial layers and excited by two shielded-loop sources. The data are in good agreement, and reveal a frequency-reduced passband demonstrating an enhancement of transmission of up to $19$dB as compared to a similarly excited vacuum-filled waveguide.
\section{Acknowledgments}
The authors would like to acknowledge the support of the Natural Sciences and Engineering Research Council (NSERC) of Canada and also the generosity of Rogers Corporation in providing donations of substrate materials.

\vspace{13pt}
\ifCLASSOPTIONcaptionsoff
  \newpage
\fi

\bibliography{BIBLIOGRAPHY}

\begin{thebibliography}{10}
\providecommand{\url}[1]{#1}
\csname url@samestyle\endcsname
\providecommand{\newblock}{\relax}
\providecommand{\bibinfo}[2]{#2}
\providecommand{\BIBentrySTDinterwordspacing}{\spaceskip=0pt\relax}
\providecommand{\BIBentryALTinterwordstretchfactor}{4}
\providecommand{\BIBentryALTinterwordspacing}{\spaceskip=\fontdimen2\font plus
\BIBentryALTinterwordstretchfactor\fontdimen3\font minus
  \fontdimen4\font\relax}
\providecommand{\BIBforeignlanguage}[2]{{%
\expandafter\ifx\csname l@#1\endcsname\relax
\typeout{** WARNING: IEEEtran.bst: No hyphenation pattern has been}%
\typeout{** loaded for the language `#1'. Using the pattern for}%
\typeout{** the default language instead.}%
\else
\language=\csname l@#1\endcsname
\fi
#2}}
\providecommand{\BIBdecl}{\relax}
\BIBdecl

\bibitem{penirschke2008microwave}
A.~Penirschke and R.~Jakoby, ``Microwave mass flow detector for particulate
  solids based on spatial filtering velocimetry,'' \emph{Microwave Theory and
  Techniques, IEEE Transactions on}, vol.~56, no.~12, pp. 3193--3199, 2008.

\bibitem{ratanadecho2002numerical}
P.~Ratanadecho, K.~Aoki, and M.~Akahori, ``A numerical and experimental
  investigation of the modeling of microwave heating for liquid layers using a
  rectangular wave guide (effects of natural convection and dielectric
  properties),'' \emph{Applied Mathematical Modelling}, vol.~26, no.~3, pp.
  449--472, 2002.

\bibitem{yin1993cyclotron}
Y.-Z. Yin, ``The cyclotron autoresonance maser with a large-orbit electron ring
  in a dielectric-loaded waveguide,'' \emph{International journal of infrared
  and millimeter waves}, vol.~14, no.~8, pp. 1587--1600, 1993.

\bibitem{nubling1996hollow}
R.~K. Nubling and J.~A. Harrington, ``Hollow-waveguide delivery systems for
  high-power, industrial {CO}$_{2}$ lasers,'' \emph{Applied optics}, vol.~35,
  no.~3, pp. 372--380, 1996.

\bibitem{clarricoats1964evanescent}
P.~J.~B. Clarricoats and B.~C. Taylor, ``Evanescent and propagating modes of
  dielectric-loaded circular waveguide,'' \emph{Proc. Inst. Elect. Eng.}, vol.
  111, pp. 1951--1956, Dec. 1964.

\bibitem{hrabar2005waveguide}
S.~Hrabar, J.~Bartolic, and Z.~Sipus, ``Waveguide miniaturization using
  uniaxial negative permeability metamaterial,'' \emph{IEEE Trans. Antennas
  Propag.}, vol.~53, no.~1, pp. 110--119, Jan. 2005.

\bibitem{meng2011controllable}
F.~Y. Meng, Q.~Wu, D.~Erni, and L.~W. Li, ``Controllable metamaterial-loaded
  waveguides supporting backward and forward waves,'' \emph{IEEE Trans.
  Antennas Propag.}, vol.~59, no.~9, pp. 3400--3411, 2011.

\bibitem{pollockmttt2013}
J.~G. Pollock and A.~K. Iyer, ``Below-cutoff propagation in metamaterial-lined
  circular waveguides,'' \emph{Microwave Theory and Techniques, IEEE
  Transactions on}, vol.~61, no.~9, pp. 3169--3178, 2013.

\bibitem{PollockISMRM2012}
J.~G. Pollock, N.~De~Zanche, and A.~K. Iyer, ``Travelling wave {MRI} at lower
  {B}$_0$ field strengths using metamaterial liners,'' \emph{Proc.
  International Society for Magnetic Resonance in Medicine}, vol.~20, p. 2792,
  2012.

\bibitem{alu2008dielectric}
A.~Alu and N.~Engheta, ``Dielectric sensing in $\epsilon$-near-zero narrow
  waveguide channels,'' \emph{Physical Review. B, Condensed Matter and
  Materials Physics}, vol.~78, no.~4, 2008.

\bibitem{duan2009research}
Z.~Duan, B.~I. Wu, S.~Xi, H.~Chen, and M.~Chen, ``Research progress in reversed
  cherenkov radiation in double-negative metamaterials,'' \emph{Progress in
  Electromagnetic Research, {PIER}}, vol.~90, pp. 75--87, 2009.

\bibitem{pollockTAP2014}
J.~G. Pollock and A.~K. Iyer, ``Miniaturized circular-waveguide probe antennas
  using metamaterial liners,'' \emph{Antennas and Propagation, IEEE
  Transactions on}, vol.~63, no.~1, pp. 428--433, Jan 2015.

\bibitem{pollockopticsx2015}
D.~Pratap, S.~A. Ramakrishna, J.~G. Pollock, and A.~K. Iyer, ``Anisotropic
  metamaterial optical fibers,'' \emph{{to appear in} Optics Express}, Mar. 19,
  2015.

\bibitem{pollockAEC2013}
J.~G. Pollock and A.~K. Iyer, ``Realization of $\epsilon$-negative-near-zero
  metamaterial liners for circular waveguides,'' \emph{4$^{th}$ Applied
  Electromagnetics Conference (AEMC 2013)}, Bhubaneswar, Orissa, India, Dec.
  2013.

\bibitem{smith2006homogenization}
D.~R. Smith and J.~B. Pendry, ``Homogenization of metamaterials by field
  averaging,'' \emph{JOSA B}, vol.~23, no.~3, pp. 391--403, 2006.

\bibitem{smith2005electromagnetic}
D.~Smith, D.~Vier, T.~Koschny, and C.~Soukoulis, ``Electromagnetic parameter
  retrieval from inhomogeneous metamaterials,'' \emph{Physical Review E},
  vol.~71, no.~3, p. 036617, 2005.

\bibitem{elser2007nonlocal}
J.~Elser, V.~A. Podolskiy, I.~Salakhutdinov, and I.~Avrutsky, ``Nonlocal
  effects in effective-medium response of nanolayered metamaterials,''
  \emph{Applied physics letters}, vol.~90, no.~19, p. 191109, 2007.

\bibitem{alu2011first}
A.~Al{\`u}, ``First-principles homogenization theory for periodic
  metamaterials,'' \emph{Physical Review B}, vol.~84, no.~7, p. 075153, 2011.

\bibitem{novotny1994light}
L.~Novotny and C.~Hafner, ``Light propagation in a cylindrical waveguide with a
  complex, metallic, dielectric function,'' \emph{Physical review E}, vol.~50,
  no.~5, p. 4094, 1994.

\bibitem{pendry1996}
J.~B. Pendry, A.~J. Holden, W.~J. Stewart, and I.~Youngs, ``Extremely low
  frequency plasmons in metallic mesostructures,'' \emph{Physical review
  letters}, vol.~76, no.~25, p. 4773, 1996.

\bibitem{demetriadou2008taming}
A.~Demetriadou and J.~Pendry, ``Taming spatial dispersion in wire
  metamaterial,'' \emph{Journal of Physics: Condensed Matter}, vol.~20, no.~29,
  p. 295222, 2008.

\bibitem{coilcraft}
Coilcraft, ``Air core inductors, 0806{SQ}6{N}0 datasheet,'' Mar. 2015.

\bibitem{whiteside1964loop}
H.~Whiteside and R.~King, ``The loop antenna as a probe,'' \emph{Antennas and
  Propagation, IEEE Transactions on}, vol.~12, no.~3, pp. 291--297, 1964.

\bibitem{enghati_tunnel}
M.~G. Silveirinha and N.~Engheta, ``Tunneling of electromagnetic energy through
  subwavelength channels and bends using $\epsilon$-near-zero materials,''
  \emph{Phys. Rev. Lett.}, vol.~97, p. 157403, 2006.

\end{thebibliography}
\bibliographystyle{IEEEtran}

\end{document}